\documentclass[a4paper,11pt]{scrartcl}

\usepackage[right=2.6cm, left=2.6cm]{geometry}
\usepackage[utf8x]{inputenc}
\usepackage{amsfonts}
\usepackage{amssymb}
\usepackage{amsmath}
\usepackage{amsthm}
\usepackage{array}
\usepackage{bbm}
\usepackage{booktabs}
\usepackage{caption}
\usepackage{color}
\usepackage{colortbl}
\usepackage{csquotes}
\usepackage{enumerate}
\usepackage{environ}
\usepackage{float}
\usepackage[acronym,toc]{glossaries}
\usepackage{graphicx}
\usepackage{hepunits}
\usepackage{ifpdf}
\usepackage{latexsym}
\usepackage{lscape}
\usepackage{makeidx}
\usepackage{mathtools}
\usepackage{multirow}
\usepackage{placeins}
\usepackage{pstricks}
\usepackage{rotating}
\usepackage{tikz}
\usepackage{units}
\usepackage{slashed}

\addtokomafont{disposition}{\boldmath} 
\KOMAoptions{bibliography=totoc}
\KOMAoptions{numbers=noendperiod}

\numberwithin{equation}{section}

\newcommand{\M}{\mathcal{M}}
\newcommand{\B}{\mathcal{B}}
\newcommand{\F}{\mathcal{F}}
\newcommand{\R}{\mathcal{R}}
\newcommand{\N}{\mathcal{N}}
\newcommand{\Op}{\mathcal{O}}
\newcommand{\diff }{{\text{d}}}
\newcommand{\Omnes}{Omn\`{e}s }
\newcommand{\MO}{Muskhelishvili--Omn\`{e}s }
\renewcommand{\vec}[1]{\mathbf{#1}}

\newcommand{\Bd}{\bar B_{d}^0}

\renewcommand{\slash}[1]{%
\mathrel{\setbox0=\hbox{$/$}\copy0\kern-\wd0\hbox{$#1$}}}
\newcommand{\slashB}{\mathrel{%
\setbox0=\hbox{$/$}\copy0\kern-1.25\wd0\hbox{$B$}}}

\newcommand{\braque}[1]{{\langle #1 \rangle}}

\newcommand{\Km}{{K^-}}
\newcommand{\Kz}{{K^0}}
\newcommand{\Kzb}{{\bar{K}^0}}
\newcommand{\pip}{\pi^+}
\newcommand{\pim}{\pi^-}
\newcommand{\piz}{\pi^0}
\newcommand{\Kbar}{\bar{K}}

\newcommand{\mpsi}{m_{\psi'}^2}

\newcommand{\bgB}{\zeta}
\newcommand{\gpsi}{\xi}

\deffootnote{1em}{1em}{\textsuperscript{\thefootnotemark}\ }
\def\be{\begin{equation}}
\def\ee{\end{equation}}

\usepackage{booktabs,array}

\newcount\rowc

\usepackage{jheppub_mod}

\title{How to employ \boldmath $\bar{B}^0_d \to J/\psi(\pi\eta,\bar{K}K)$ decays to extract information on $\pi\eta$ scattering}

\author[a]{M.~Albaladejo,\footnote{Current address: Departamento de F\'isica, Facultad de Qu\'imica, Universidad de Murcia, E-30071, Murcia, Spain; e-mail: albaladejo@um.es.}}
\author[b]{J.~T.~Daub,}
\author[c]{C.~Hanhart,}
\author[b]{B.~Kubis,}
\author[d]{and B.~Moussallam}

\affiliation[a]{
Instituto de F\'isica Corpuscular (IFIC), Centro Mixto CSIC-Universidad de Valencia,\\
Institutos de Investigaci\'on de Paterna, Aptdo. 22085, 46071 Valencia, Spain}

\affiliation[b]{
Helmholtz-Institut f\"ur Strahlen- und Kernphysik (Theorie) and \\ 
Bethe Center for Theoretical Physics, Universit\"at Bonn, 
53115 Bonn, Germany}

\affiliation[c]{
Institut f\"ur Kernphysik, Institute for Advanced Simulation, 
and J\"ulich Center for Hadron Physics,
Forschungszentrum J\"ulich, 52425 J\"{u}lich, Germany}

\affiliation[d]{
Groupe de Physique Th\'eorique IPN (UMR8608),
Universit\'e Paris-Sud 11, 91406 Orsay, France}

\emailAdd{Miguel.Albaladejo@ific.uv.es}
\emailAdd{daub@hiskp.uni-bonn.de}
\emailAdd{c.hanhart@fz-juelich.de}
\emailAdd{kubis@hiskp.uni-bonn.de}
\emailAdd{moussall@ipno.in2p3.fr}

\abstract{
We demonstrate that dispersion theory allows one to deduce crucial information
on $\pi\eta$ scattering from the final-state interactions of the light mesons
visible  in the spectral distributions of the
decays $\bar{B}^0_d \to J/\psi(\pi^0\eta,K^+K^-,K^0\bar K^0)$. Thus high-quality measurements
of these differential observables are highly desired. The corresponding rates are
predicted to be of the same order of magnitude as those for $\bar{B}^0_d \to J/\psi\pi^+\pi^-$
measured recently at LHCb, letting the corresponding measurement appear feasible.
}
\begin{document}
\maketitle

\section{Introduction}

The interactions of the lightest hadrons, the pions, with themselves as well as with kaons, the next-lightest
strongly-interacting particles within the pseudoscalar ground-state octet, are known to excellent precision.
The combination of dispersion relations in the form of Roy or Roy--Steiner equations, constrained by 
chiral perturbation theory at lowest energies and using experimental data as input, has increased our knowledge
of the leading partial waves of pion--pion~\cite{Martin:1976mb,Ananthanarayan:2000ht,Colangelo:2001df,GarciaMartin:2011cn}
and pion--kaon~\cite{BDM04} scattering enormously.
This has a large impact on a wide range of scattering or decay processes in which pions and kaons are produced:
dispersion relations allow to relate the final-state interactions to the scattering phase shifts in a model-independent
way~\cite{Barton:1965,ElBennich:2009da,GarciaMartin:2010cw,Hoferichter:2011wk,Stollenwerk:2011zz,Hoferichter:2012pm,Kang:2013jaa,Dai:2014zta,Colangelo:2015kha,Kubis:2015sga,Daub:2015xja,Niecknig:2015ija}.
Both reactions---$\pi\pi$ and $\pi K$ scattering---are furthermore closely intertwined in the isospin-0 $S$-wave
system, where the reaction $\pi\pi\to \bar KK$, the crossed process of $\pi K$ scattering,
dominates the inelasticity in pion--pion scattering near the 
$\bar KK$ threshold, the region of the $f_0(980)$ resonance;
for these (scalar) quantum numbers, a coupled-channel treatment is mandatory~\cite{DGL90,Moussallam2000}.

Information on the scattering of pions off the $\eta$, which would complete our understanding of 
pion reactions off the pseudoscalar ground-state octet, is much scarcer. 
Two important resonances are known in the $I(J^P) = 1(0^+)$ sector, namely the $a_0(980)$ and $a_0(1450)$~\cite{Olive:2016xmw}. 
Several models for the $\pi\eta$ $S$-wave scattering amplitude have been proposed in the literature~\cite{Flatte:1976xu,Oller:1998hw,Black:1999dx,GomezNicola:2001as,Furman:2002cg}, some of them constrained by the results from chiral perturbation
theory at threshold~\cite{Bernard:1991xb}.
Very recently, first information about this amplitude has also come from lattice QCD simulations~\cite{Dudek:2016cru,Guo:2016zep}.
There is one remarkable similarity of $\pi\eta$ $S$-wave scattering to the $\pi\pi$ $I=0$ $S$-wave, in that
also for $\pi\eta$, the first important inelastic channel is given by $\bar KK$, whose threshold also there
coincides with the presence of a scalar resonance, the isospin-1 $a_0(980)$ resonance.  
Therefore also in this case, a coupled-channel treatment of $\pi\eta$ and $\bar KK$ (in $I=1$) is required
to describe the energy region around $1\,\GeV$.
A corresponding unitary $T$-matrix has recently been constructed in ref.~\cite{Albaladejo:2015aca}, 
to which chiral constraints~\cite{Gasser:1984gg} have been imposed 
as well as experimental information available on the $a_0(980)$ and $a_0(1450)$ resonances.
However, the result still has considerable uncertainties due to the limited accuracy of the experimental input.

In this article, we give a theoretical prediction for the decay $\bar{B}_d^0 \to J/\psi \pi^0 \eta$, not measured so far, in which the properties of the $I=1$ $\pi\eta$--$\bar{K}K$ $S$-wave system and the $a_0(980)$ resonance could be further tested. This decay is well suited for such a purpose, because the production mechanism is rather strongly constrained. It is closely related to $\Bd \to J/\psi \pi^+ \pi^-$, which has been analyzed in ref.~\cite{Daub:2015xja}
(together with the similar $\bar B^0_s \to J/\psi \pi^+ \pi^-$).
There it has been demonstrated that the $S$-waves can be described very well in terms of pion (and kaon) scalar form factors, 
constructed in a \MO formalism. Due to the $\pi\pi$--$\bar KK$ channel coupling, the isoscalar $\bar KK$ $S$-wave component is related unambiguously 
to the $\pi\pi$ final state.
However, as the source term is given dominantly in terms of $\bar dd$ quark bilinears, it produces isoscalar and
isovector meson pairs with known relative sign and strengths. With the strength of the dimeson source 
fixed from experimental data on $\Bd \to J/\psi \pi^+ \pi^-$~\cite{Daub:2015xja}, 
and given the analogous isovector scalar form factors for $\pi\eta$ and $\bar KK$~\cite{Albaladejo:2015aca},
we can thus give an absolute prediction for the $J/\psi \pi^0 \eta$ channel.  
The combination of isoscalar and isovector scalar form factors for the kaons then allows one to fully analyze the 
physical $K^+K^-$ and $K^0\bar{K}^0$ final states.
A measurement of these systems (or likewise the pertinent decay of the charged $B$-meson associated with the scalar 
isovector current $\bar u d$) therefore would allow to further constrain theoretical models for $\pi\eta$
scattering, its phase shifts and inelasticities.
Similar predictions for the decays of the $\bar B^0_{d,s}$ and $B^-$ mesons into $J/\psi$ and isoscalar and isovector meson pairs have been discussed in the chiral unitary approach in ref.~\cite{Liang:2015qva} (see also ref.~\cite{Oset:2016lyh} for a general review about the use of the chiral unitary approach to study the final-state strong interactions in weak decays).

The outline of this article is as follows. In section~\ref{sec:kinematics}, elementary kinematics of the decay $\Bd \to J/\psi \pi^0 \eta$ as well as the calculation of decay rates in terms of the corresponding matrix elements are described. Flavor relations based on chiral symmetry are discussed in section~\ref{sec:chiralsection}.  We introduce the required dimeson scalar form factors in section~\ref{sec:omnes}, and briefly recapitulate the \Omnes formalism used to describe the latter. Results for $\bar{B}^0_d \to J/\psi \left\{\pi\eta,\bar{K}K\right\}$ are presented in section~\ref{sec:resultsB0d}. We conclude in section~\ref{sec:summary}.  Two appendices provide further arguments for our treatment of the $S$-waves in terms of scalar form factors, as well as their dominance for the decays under investigation:  appendix~\ref{app:chiral} discusses the relation between scalar form factors and the contributions of $S$-waves in these $B$-decays by means of a chiral Lagrangian; and in appendix~\ref{app:Pwaves}, we demonstrate the suppression of $\pi\eta$ $P$-wave contributions both in terms of generic (chiral) power counting arguments, as well as by two explicit calculations of $\psi(2S)$ and $B^*$ exchanges that simultaneously address the importance of left-hand cuts.

\section{Kinematics, form factors, decay rate}\label{sec:kinematics}

\begin{figure}
\centering  
\includegraphics[scale=0.6]{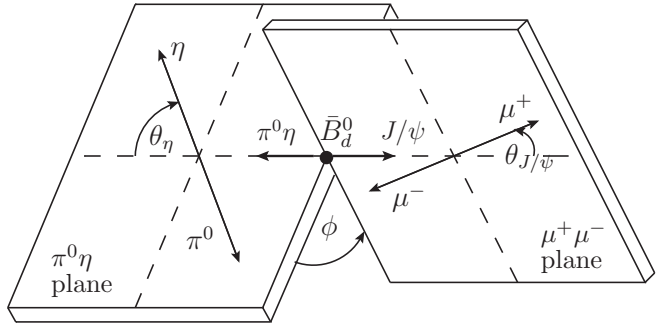}
\caption{Definition of the kinematical variables for $\Bd \to J/\psi \pi^0\eta $.}
\label{fig:kinematics}
\end{figure}

The kinematics of the decay $ \Bd (p_B) \to J/\psi (p_\psi) \pi^0 (p_1) \eta (p_2) $ ($J/\psi \to \mu^+ \mu^-$) can be described by four variables: 
the invariant dimeson mass squared, $s=(p_1 + p_2)^2$, 
and three helicity angles, see figure~\ref{fig:kinematics}, namely
\begin{itemize}  
\item the angle $\theta_{J/\psi}$ between the $\mu^+$ in the $J/\psi$ rest frame ($\Sigma_{J/\psi}$) and the $J/\psi$ in the $\Bd$ rest frame ($\Sigma_B$);
\item the angle $\theta_\eta$ between the $\eta$ in the $\pi^0\eta$ center-of-mass frame $\Sigma_{\pi\eta}$ and the $\pi^0\eta$ line-of-flight in $\Sigma_B$;
\item the angle $\phi$ between the $\pi^0\eta$ and the dimuon planes, where the latter originate from the decay of the $J/\psi$.
\end{itemize}
The three-momenta of the pion or the $\eta$ in the $\pi^0\eta$ center-of-mass system ($\vec p_{\pi\eta}$) and that of the $J/\psi$ in the $\Bd$ rest frame ($\vec p_\psi$) are given by 
\be 
|\vec p_{\pi\eta} |=\frac{\lambda^{1/2}(s,M_\pi^2,M_\eta^2)}{2\sqrt{s}} \equiv \frac{Y\sqrt{s}}{2}, \quad
|\vec p_\psi |=\frac{\lambda^{1/2}(s,m_\psi^2,m_B^2)}{2 m_B} \equiv \frac{X}{m_B},
\label{eq:kindef}
\ee
with the  K\"all\'{e}n function $\lambda(a,b,c) = a^2+b^2+c^2 - 2(ab +ac +bc)$.
We define two further Mandelstam variables as
\be
t = (p_B - p_1)^2 \quad \mathrm{and} \quad u = (p_B - p_2)^2,
\ee
where the difference of these two determines the scattering angle $\theta_\eta$,
\be
t - u 
= - 2 Y X \cos \theta_\eta + \frac{\Delta_m}{s}, \quad \Delta_m = (m_B^2-m_\psi^2)(M_\eta^2-M_\pi^2).
\ee
We do not spell out the kinematic relations for the $\bar KK$ final state explicitly---they can be obtained from the 
above in a straightforward manner.

We decompose the matrix element for the decay in the following way:
\begin{align}
\M_{fi} &= \frac{G_F}{\sqrt{2}} V_{cb} V_{cd}^*\, \M^{\mathrm{eff}} , \nonumber\\
\M^{\mathrm{eff}} &= f_\psi m_\psi \epsilon_\mu^* (p_\psi,\lambda)
\left( \frac{m_\psi P_{(0)}^{\mu}}{X}  \F_0 + \frac{Q_{(\parallel)}^\mu}{\sqrt{s}}\F_\parallel - \frac{i \bar p_{(\perp)}^\mu}{\sqrt{s}} \F_\perp \right) . \label{eq:matrixelement}
\end{align}
$f_\psi \approx 405\,\MeV$ denotes the decay constant of the $J/\psi$, and $\epsilon_\mu^* (p_\psi,\lambda)$ the corresponding polarization vector
of helicity $\lambda$.
The three transversity form factors $\F_0$, $\F_\parallel$, and $\F_\perp$ correspond to the orthogonal basis of momentum vectors
\begin{align}\label{eq:newbasis}
P_{(0)}^\mu &= P^\mu - \frac{P\cdot p_\psi}{p_\psi^2} p_\psi^\mu, \qquad
\bar p_{(\perp)}^\mu = \frac{\epsilon^{\mu\alpha\beta\gamma}}{X} (p_\psi)_\alpha P_\beta Q_\gamma, \nonumber\\
Q_{(\parallel)}^\mu &= Q^\mu +\frac{p_\psi^2(P\cdot Q)-(P \cdot p_\psi)(Q \cdot p_\psi)}{X^2} P^\mu + \frac{P^2 (Q \cdot p_\psi)-(P\cdot p_\psi)(P\cdot Q)}{X^2} p_\psi^\mu,
\end{align}
where $P^\mu = p_1^\mu+p_2^\mu$ and $Q^\mu = p_1^\mu-p_2^\mu$, thus 
generalizing the formulae discussed previously~\cite{Faller:2013dwa,Daub:2015xja} to unequal masses.
Again, the relations for the $\bar KK$ final state simplify accordingly.
We employ $\epsilon_{\mu\alpha\beta\gamma}$ with the convention $\epsilon_{0123} = - \epsilon^{0123} = +1$.
The partial-wave expansions of the transversity form factors up to $D$-waves read
\begin{align} 
\F_0 (s,\theta_\eta) 
&= \F_0^{(S)} (s) + \sqrt{3} \cos\theta_\eta \F_0^{(P)} (s)  + \frac{\sqrt{5}}{2} \left(3 \cos^2\theta_\eta - 1 \right) \F_0^{(D)} (s) + \dots \,, \nonumber\\
\F_{\parallel,\perp} (s,\theta_\eta) &= \sqrt{\frac{3}{2}} \left( \F_{\parallel,\perp}^{(P)}(s) + \sqrt{5} \cos \theta_\eta \F_{\parallel,\perp}^{(D)}(s) \right) + \dots. \label{eq:F-PWE}
\end{align}
Note that the partial waves defined this way still contain kinematical zeros that have to be
removed for a dispersive treatment~\cite{Daub:2015xja}.

The structure of the matrix element in eq.~\eqref{eq:matrixelement}, its decomposition in terms of transversity form factors, 
as well as the partial-wave expansion of the latter is independent of any factorization assumption, and entirely general.
Factorization of the form
\begin{align}
\M^{\mathrm{eff}} = a^{\mathrm{eff}} (\mu) \M^{\pi\eta}_\mu \M^{c \bar c\, \mu} + \ldots, \qquad
\M^{\pi\eta}_\mu &= \langle \pi^0(p_1) \eta(p_2) | \bar{d} \gamma_\mu (1-\gamma_5) b | \Bd (p_B) \rangle , \nonumber\\
\M^{c \bar c\, \mu} &= \langle J/\psi(p_\psi, \epsilon) | \bar{c} \gamma^\mu c | 0 \rangle , \label{eq:factorization}
\end{align}
which leads to the same decomposition as in eq.~\eqref{eq:matrixelement}, is only required if we want to identify the transversity
form factors explicitly with those that, after an isospin rotation, describe the semileptonic decays 
$\Bd \to \pi^+\eta \ell^- \bar\nu_\ell$; no such attempt is made in the following study.
The effective coupling $a^{\mathrm{eff}} (\mu)$ is related to Wilson coefficients of the effective Hamiltonian for the $b\to c\bar cd$ transition~\cite{Colangelo:2010wg,Daub:2015xja}; the ellipsis in eq.~\eqref{eq:factorization} denotes higher-order corrections to factorization that compensate for the scale dependence in $a^{\mathrm{eff}} (\mu)$~\cite{Beneke:2000ry}.

The differential decay rate for the $\Bd \to J/\psi \pi^0 \eta$ decay is given by
\begin{align}
&\frac{\diff^2\Gamma}{\diff\sqrt{s} \,\diff\cos\theta_\eta} \nonumber\\
& =  \frac{G_F^2 |V_{cb}|^2 |V_{cd}|^2 f_\psi^2 m_\psi^2 X Y \sqrt{s}}{4 (4\pi)^3 m_B^3} \, 
\Big\{        \vert \F_0 (s,\theta_\eta) \vert ^2
    + Y^2 \sin^2\theta_\eta \left(
                 \vert \F_\parallel(s,\theta_\eta) \vert ^2
                +\vert \F_\perp(s,\theta_\eta) \vert ^2\right) \Big\} \nonumber\\
& \approx 
\frac{G_F^2 |V_{cb}|^2 |V_{cd}|^2 f_\psi^2 m_\psi^2 X Y \sqrt{s}}{4 (4\pi)^3 m_B^3}  \nonumber\\
& \quad \times\bigg\{ \Big|\F_0^{(S)} (s) + \sqrt{3} \cos\theta_\eta \F_0^{(P)} (s)  + \frac{\sqrt{5}}{2} \left(3 \cos^2\theta_\eta - 1 \right) \F_0^{(D)} (s)  \Big|^2 \nonumber\\
& \qquad+\frac{3}{2} Y^2 \sin^2 \theta_\eta\bigg(\Big|\F_\parallel^{(P)} + \sqrt{5} \cos \theta_\eta \F_{\parallel}^{(D)}(s) \Big|^2 +  \Big|\F_\perp^{(P)} + \sqrt{5} \cos \theta_\eta \F_{\perp}^{(D)}(s) \Big|^2 \bigg) \bigg\}, \label{eq:d2Gamma}
\end{align}
where in the second step partial waves up to and including $D$-waves are considered.  Most of the discussion in this article will concentrate on the $S$-wave only.
We show in appendix~\ref{app:Pwaves} that the $P$-wave is very small:
as we discuss in appendix~\ref{app:Pwave-chiral}, the production vertex for a $\pi\eta$ $P$-wave is chirally suppressed;
it is generically smaller relative to the $S$-wave by a factor of $Y M_K^2/(4\pi F_\pi)^2$, which around $1\,\GeV$ 
amounts to one order of magnitude.  In appendices~\ref{app:psi2S} and \ref{app:B*}, 
we have calculated the $P$-wave contributions generated by $t$- and $u$-channel resonance exchanges, and found
them to be even more suppressed than this generic estimate.
Furthermore, the $\pi\eta$ $P$-wave has exotic quantum numbers;
final-state-interaction effects should thus be small below $1\,\GeV$.
We therefore find it safe to assume it to be negligible in the energy
range considered in this article, and well within the uncertainty of the $S$-wave contribution.
We will briefly discuss the impact of $D$-waves in section~\ref{sec:pieta}, mainly to demonstrate that they also only
become important for dimeson energies well above $1\,\GeV$.

Angular integration of eq.~\eqref{eq:d2Gamma} yields the differential decay rate
\begin{align}\label{eq:Y00etapi}
\frac{\diff\Gamma}{\diff\sqrt{s}} &= \frac{X Y \sqrt{s}}{2 m_B} \,\big|\,\mathcal{C} \big|^2
\bigg\{ \Big|\F_0^{(S)} (s)\Big|^2 + \Big| \F_0^{(P)} (s)\Big|^2  + \Big| \F_0^{(D)} (s)\Big|^2 \nonumber\\
&\quad + Y^2 \bigg(\Big|\F_\parallel^{(P)}(s)\Big|^2 + \Big| \F_{\parallel}^{(D)}(s) \Big|^2 +  \Big|\F_\perp^{(P)}(s)\Big|^2 + \Big| \F_{\perp}^{(D)}(s) \Big|^2\bigg) \bigg\} \nonumber\\
&= \sqrt{4 \pi}  \langle Y _0 ^0 \rangle (s), \qquad 
\mathcal{C}=\frac{G_F V_{cb} V_{cd}^* f_\psi m_\psi }{\sqrt{(4\pi)^3}\, m_B} ,
\end{align}
where the second equality relates to the angular moment $\langle Y _0 ^0 \rangle (s)$ commonly used by the LHCb collaboration, in ref.~\cite{Daub:2015xja} as well as later in this text.

\section{Chiral-symmetry-based relations}\label{sec:chiralsection}
We discuss now how a relation can be derived between the amplitudes 
$\Bd \to J/\psi \pi^+\pi^-$ and $\Bd \to J/\psi \pi^0\eta$ based on chiral
symmetry. We start by writing the effective weak Hamiltonian in terms of the
usual set of 14 operators,
\be\label{eq:Heff}
{\cal H}_{\text{eff}}= \frac{G_F}{\sqrt2}\sum_{p=u,c} \lambda_p\Big(
C_1 Q^p_1 + C_2 Q^p_2 +\sum_{i=3,\ldots, 10} C_i Q_i
+C_{7\gamma} Q_{7\gamma}  +C_{8g} Q_{8g}
\Big) + \text{h.c.},
\ee
using the same notation as in ref.~\cite{Beneke:2001ev} except for the obvious
replacement of $s$ by $d$, in particular 
\be
\lambda_c= V_{cb}^{\phantom{*}} V_{cd}^*, \qquad
\lambda_u= V_{ub}^{\phantom{*}} V_{ud}^*,
\ee
and
\begin{align}
Q_1^c &= 4\,\bar{c}_L \gamma^\mu b_L\, \bar{d}_L \gamma_\mu c_L  , &
Q_2^c &=  4\,\bar{c}_L^i \gamma^\mu b_L^j\,  \bar{d}_L^j \gamma_\mu c_L^i , \nonumber\\
Q_1^u &= 4\,\bar{u}_L \gamma^\mu b_L \, \bar{d}_L \gamma_\mu u_L  , &
Q_2^u &=  4\,\bar{u}_L^i \gamma^\mu b_L^j \, \bar{d}_L^j \gamma_\mu u_L^i ,
\end{align}
where $q_{L,R}=\frac{1}{2}(1\mp\gamma_5) q$. For the processes under consideration it
is clear, at first, that the electromagnetic penguin operators $Q_{7-10}$ as well as $Q_{7\gamma}$ can be neglected compared to $Q_1^c$,
$Q_2^c$. We make the further assumption that the two operators $Q_1^u$,
$Q_2^u$ can also be neglected. This is justified by the OZI rule: in order to
produce a $J/\psi$ in the final state from the operators $Q_1^u$, $Q_2^u$ one
must proceed via quark-disconnected diagrams involving three gluons. The OZI
rule is known to be quite effective for heavy-quarkonium production or decays. 

The remaining seven operators all transform simply as $\bar{d}_L$ under the
$SU(3)_L\times SU(3)_R$ chiral group. We can then construct a chiral
Lagrangian that encodes this transformation property and describes the
dynamics in the region where the light pseudoscalar meson pair has a very
small energy; this is detailed in appendix~\ref{app:chiral}. At leading order in the chiral
expansion, the following relation can be derived between the $\Bd$-decay
amplitudes and the vacuum matrix elements of $\bar{d}d$ operators:
\begin{equation} \label{eq:symmetryrelation}
\frac{\langle J/\psi (\pi^0\eta)_{l=0} | \Op_{\bar d} | \Bd \rangle}
{\langle J/\psi (\pi^+\pi^-)_{l=0} | \Op_{\bar d} | \Bd \rangle} 
= \frac{\langle \pi^0\eta | \bar{d}d | 0 \rangle}{\langle \pi^+\pi^- | \bar{d}d | 0 \rangle}.
\end{equation}
This relation will be used at energy $s=0$.

We emphasize that it is important to use chiral symmetry to derive this result rather than flavor
symmetry alone. Indeed, under the flavor symmetry group $SU(3)_F$, the state
$(\pi\pi)_{I=0}$ appears both in the singlet and in the octet representations,
while $(\pi\eta)$ belongs to the octet. Therefore, no relation can be derived
between the corresponding $\Bd$ decay amplitudes based on flavor symmetry alone.
For simplicity, we will however still refer to these as flavor relations in the following.

\section{Scalar form factors and Omn\`es formalism}\label{sec:omnes}

We treat the $\Bd \to J/\psi \pi^0\eta$ decay in an analogous manner to the $J/\psi \pi^+\pi^-$ final state in ref.~\cite{Daub:2015xja}, the isospin-0 counterpart of the process considered here. In that case, experiment has shown that there is no structure visible in the (exotic) $J/\psi \pi^+$ channel, thus any crossed-channel processes should be negligible: 
there are no significant left-hand cuts in the decay amplitude, which therefore could be described in terms
of form factors containing a right-hand cut only. 
This assumption is adopted in the ongoing study, where, however, the justification is not quite so clear:
e.g.\ the $\psi(2S)$ decays into $J/\psi \eta$, and therefore will show up as a resonance in the corresponding
distribution.\footnote{Actually, the same problem arises already for $J/\psi \pi^0$, another observed
decay channel of the $\psi(2S)$, which however breaks isospin symmetry and hence is very weak.}
However, due to the anomalous nature of the $\psi(2S)\to J/\psi\eta$ vertex, this $\psi(2S)$-exchange 
mechanism only contributes to the transversity form factor $\F_\perp$,
whose partial-wave expansion begins with a $P$-wave, see eq.~\eqref{eq:F-PWE}, and hence cannot contribute to the 
$\pi^0\eta$ $S$-wave that we will concentrate on below.
In addition, the coupling $\psi(2S) \to J/\psi \eta$ violates the OZI rule and is still rather weak: 
compare 
\begin{align}
\B(\Bd \to \psi(2S) \pi^0) &= (1.17 \pm 0.19) \times 10^{-5}~\text{\cite{Chobanova:2015ssy}}, \nonumber\\
\B(\psi(2S) \to J/\psi \,\eta) &= (3.36 \pm 0.05) \times 10^{-2}~\text{\cite{Olive:2016xmw}} 
\label{eq:psi2SBR}
\end{align}
to
$\B(\Bd \to J/\psi \pi^+\pi^-)  = (4.03 \pm 0.18) \times 10^{-5} $~\cite{Olive:2016xmw};
we will see below that the branching ratio $\B(\Bd \to J/\psi \pi^0\eta)$ is predicted to be of a comparable size.
In appendix~\ref{app:psi2S}, we calculated the $\psi(2S)$-exchange contribution explicitly and showed that 
below $\sqrt{s} \approx 1.34\,\GeV$, where the $\psi(2S)$ cannot go on-shell, its effect is even more suppressed.
We assume that other charmonium resonances whose exchange in the $t$- (or even $u$-)channel 
\textit{can} contribute to the $\pi^0\eta$ $S$-wave
(such as axialvector ones of negative $C$-parity) couple similarly weakly.
Furthermore, in appendix~\ref{app:B*} we have studied the effect of $t$- and $u$-channel $B^*$-exchange diagrams, 
whose cut contributions lie well outside the physical decay region; however even then, the pole terms are 
suppressed in the $S$-wave due to chiral and heavy-quark symmetry.  
We therefore neglect the influence of left-hand cuts altogether.

Adopting the observation of ref.~\cite{Daub:2015xja} that the $\Bd \to J/\psi \pi^+\pi^-$ $S$-wave is indeed proportional to the scalar form factor (in particular, not even a linear polynomial is required at the present accuracy of the data~\cite{Aaij:2014siy}), the $\Bd \to J/\psi \pi^0\eta$ $S$-wave amplitude will by analogy be proportional to the scalar $\pi\eta$ isovector form factor defined via
\be\label{eq:pietaff}
\left\langle \pi^0(p_1) \eta(p_2) \left|\tfrac{1}{2}(\bar u u - \bar d d)\right|0\right\rangle = \B_0\Gamma_{\pi\eta}^{I=1}(s),
\ee
which is calculated in a coupled-channel formalism in ref.~\cite{Albaladejo:2015aca}.
Both the isoscalar and the isovector meson pairs are generated from a pure $\bar d d$ source.
The isospin decomposition 
of the scalar current reads
\be\label{eq:dd}
\bar d d = -\frac{1}{2} \big( \bar u u - \bar d d\big)
+ \frac{1}{2} \big( \bar u u + \bar d d\big),
\ee
from which we read off the relative strength of the isoscalar to the isovector component, $\eta_0/\eta_1 = -1$.
Thus given a known isoscalar $S$-wave $\mathcal{C} \F_0^{(S,I=0)}(s)= X \bar b_0^n \Gamma_{\pi\pi}^{I=0}(s)$
(where $\bar b_0^n$ is a subtraction/normalization constant related to the fit constant $b_0^n$ introduced in ref.~\cite{Daub:2015xja} and the kinematic
variable $X$ was defined in eq.~\eqref{eq:kindef}), 
with the scalar isoscalar pion form factor 
\be
\B_0\Gamma_{\pi\pi}^{I=0} (s) =\left\langle \pi^+(p_1) \pi^-(p_2) \left|\tfrac{1}{2}(\bar u u + \bar d d)\right|0\right\rangle ,
\ee
we can predict the isovector $S$-wave
\begin{equation}\label{eq:F0s}
\mathcal{C} \F_0^{(S,I=1)}(s)= - X \bar b_0^n \Gamma_{\pi\eta}^{I=1}(s).
\end{equation}
The constant $\bar b_0^n$ is the same as for the $\pi^+\pi^-$ final state precisely due to the symmetry relation~\eqref{eq:symmetryrelation}.
Similarly, potential linear terms in $s$ multiplying the scalar form factors would also be symmetry-related for both meson pairs
under consideration: as long as the data do not suggest the necessity to include such a term in the description of the 
$\Bd\to J/\psi \pi^+\pi^-$ $S$-wave, it will be negligible also in $\Bd\to J/\psi \pi^0\eta$. 
Adopting the fit results of ref.~\cite{Daub:2015xja} we have to take into account that the experimental data for the $\Bd \to J/\psi \pi^+\pi^-$ spectrum~\cite{Aaij:2014siy} (used for the determination of the parameter $b_n^0$) is normalized arbitrarily, i.e.\ we have to achieve a proper normalization of the $\pi^+\pi^-$ distribution that we then adapt to the $\pi\eta$ prediction. For this purpose we use the absolute branching fraction
$
\B\left(\bar B_{d}^0 \to J/\psi \pi^+\pi^-\right) = (4.03 \pm 0.18) \times 10^{-5} 
$~\cite{Olive:2016xmw}
and define the strength $\bar b_0^n = \sqrt{\N_\pi} b_0^n$ with the normalization constant~\cite{Daub:2015xja}
\be
\N_{\pi} = \frac{\B(\bar B_{d}^0 \to J/\psi \pi^+\pi^-)\, \Gamma^{\mathrm{tot}} (\bar B^0_d)}{N(\bar B_{d}^0 \to J/\psi \pi^+\pi^-)}  ,
\ee
where $N(\bar B_{d}^0 \to J/\psi \pi^+\pi^-) =  24080.5 \pm 148$ is the total number of signal events in ref.~\cite{Aaij:2014siy} and $\Gamma^{\mathrm{tot}} (\bar B^0_d) = 1/\tau (\bar B^0_d) $, $\tau (\bar B^0_d) = (1.519 \pm 0.005)  \cdot 10^{-12} s$~\cite{Olive:2016xmw}.
Two different values were obtained for $b_0^n$, namely (a) $b_0^n = 10.3\,\GeV^{-7/2}$ corresponding to a fit with only constant subtraction polynomials for $S$- and $P$-waves, and (b) $b_0^n=10.6\,\GeV^{-7/2}$, corresponding to a fit with a linear (in $s$) contribution to the $P$-wave polynomial as well as a $D$-wave contribution, which had an effect on the $\pi\pi$ $S$-wave~\cite{Daub:2015xja,Aaij:2014siy}.
We therefore obtain $\bar b_0^n = 2.77 \cdot 10^{-10}~\GeV^{-3}$ and $\bar b_0^n = 2.85 \cdot 10^{-10}~\GeV^{-3}$, respectively.
Due to the large uncertainty arising from the input phase $\delta_{12}$ (see figure~\ref{fig:gammapieta}) the error in the normalization constant is completely negligible. 

By means of an isospin rotation the $\pi\eta$ form factor eq.~\eqref{eq:pietaff} likewise describes the transition to a charged $\pi^+ \eta$ pair via a $\bar u d$ source,
\be
\left\langle \pi^+(p_1) \eta(p_2) \left| \bar u d\right|0\right\rangle = \sqrt{2}\B_0\Gamma_{\pi\eta}^{I=1}(s).
\ee
This straightforwardly allows for a prediction of the charged $B^\pm \to J/\psi \pi^\pm \eta$ mode as well, whose differential decay rate $\diff\Gamma/\diff\sqrt{s}$ differs from the $\bar B^0_d$ one, eq.~\eqref{eq:Y00etapi}, by a factor 2 (except for negligible kinematical replacements due to isospin-violating mass differences).

In a similar way the $\Bd \to J/\psi K^+ K^- /  K^0 \bar K^0$ angular moments are derived. Aside from the appropriate kinematical replacements,
now both the isoscalar and isovector $S$-wave components contribute, with known relative strengths $\eta_0/\eta_1 = -1$ or $+1$ for charged or neutral kaon systems, respectively, i.e.\ the $S$-wave amplitude reads
\begin{align}
\mathcal{C} \mathcal{G}_c^{(S)}(s) & =  X \bar b_0^n \big(\Gamma_{KK}^{I=0}(s)-\Gamma_{KK}^{I=1}(s)\big), \nonumber \\
\mathcal{C} \mathcal{G}_n^{(S)}(s) & =  X \bar b_0^n \big(\Gamma_{KK}^{I=0}(s)+\Gamma_{KK}^{I=1}(s)\big), \label{eq:G0s}
\end{align}
where the kaon form factors are defined in analogy to the $\pi\pi$ and $\pi\eta$ ones. 
  
The strong coupling of two $S$-wave pions in the isoscalar case and the $\pi \eta$ pair in the isovector one to $\bar KK$ near $1\,\GeV$ due to the $f_0(980)$ and $a_0(980)$ resonances, respectively, causes a sharp onset of the $\bar KK$ inelasticity. This necessitates a coupled-channel treatment of the scalar isoscalar and isovector form factors and we solve the $I=0$ $\pi\pi$--$\bar K K$ and the $I=1$ $\pi\eta$--$\bar K K$ two-channel \MO problems.

For the isoscalar solution of the \MO problem the $T$-matrix parametri\-zation requires \textit{three} input functions: in addition to the $\pi\pi$ phase shift already necessary in the 
elastic case, modulus and phase of the $\pi\pi \to \bar{K}K$ $S$-wave amplitude also need to be known. 
For our main solution, the $\pi\pi$ phase shift is obtained from the Roy equation analysis by the Bern group~\cite{CCL2012},
the modulus of the $\pi\pi \to \bar{K}K$ $S$-wave is known from the solution of Roy--Steiner equations for 
$\pi K$ scattering performed in Orsay~\cite{BDM04}, and its phase from partial-wave analyses~\cite{Cohen, Etkin}.
In order to assess the uncertainty of these parametrizations, as an alternative
we also employ the coupled-channel $T$-matrix constructed in ref.~\cite{Dai:2014zta} 
by fitting to $\pi\pi$ data, the Madrid--Krak\'ow Roy-equation analysis~\cite{GarciaMartin:2011cn} at low energies,
as well as to Dalitz plots of $D_s^+\to \pi^+\pi^-\pi^+$~\cite{Aubert:2008ao} and $D_s^+\to K^+K^-\pi^+$~\cite{delAmoSanchez:2010yp}.

For the isovector sector, we use the approach of ref.~\cite{Albaladejo:2015aca}. In that work, a coupled-channel $T$-matrix is constructed that fulfills unitarity, and the amplitudes are approximately matched with the perturbative ones derived from $\mathcal{O}(p^4)$ chiral perturbation theory. With this method six phenomenological parameters are introduced, to be determined by experimental information about the $a_0(980)$ and $a_0(1450)$ resonances. Specifically, five experimental constraints are imposed, and hence there is still a one-parameter freedom in the model that can be associated with the sum of the phase shifts of the $\pi\eta$ ($\delta_{11}$) and $\bar{K}K$ ($\delta_{22}$) 
channels at the mass of the $a_0(1450)$,
$\delta_{12} \equiv (\delta_{11} + \delta_{22})({s}=m^2_{a_0(1450)})$.

The scalar coupled-channel form factors in the \Omnes formalism read 
\be
\left(
\begin{array}{c}
\Gamma_{\pi\pi}^{I=0} (s) \\ \frac{2}{\sqrt{3}} \Gamma_{KK}^{I=0}(s) 
\end{array}
\right)
=
\left(
\begin{array}{cc}
\Omega_{11}^{I=0} (s) & \Omega_{12}^{I=0} (s)\\ \Omega_{21}^{I=0}(s) & \Omega_{22}^{I=0} (s)
\end{array}
\right) \cdot
\left(
\begin{array}{c}
\Gamma^{I=0}_{\pi\pi} (0) \\ \frac{2}{\sqrt{3}} \Gamma^{I=0}_{KK}(0) 
\end{array}
\right)
\ee
for the isoscalar meson pairs, and
\be
\left(
\begin{array}{c}
\Gamma^{I=1}_{\pi\eta} (s) \\ \sqrt{2} \Gamma^{I=1}_{KK}(s) 
\end{array}
\right)
=
\left(
\begin{array}{cc}
\Omega_{11}^{I=1} (s) & \Omega_{12}^{I=1} (s)\\ \Omega_{21}^{I=1}(s) & \Omega_{22}^{I=1} (s)
\end{array}
\right) \cdot
\left(
\begin{array}{c}
\Gamma^{I=1}_{\pi\eta} (0) \\ \sqrt{2} \Gamma^{I=1}_{KK}(0) 
\end{array}
\right)
\ee
for the isovector $\pi \eta$--$\bar K K$ system.
The resulting form factors depend on two normalization constants $\Gamma_{PP'}^I(0)$ each that we constrain from their chiral one-loop (or next-to-leading-order) representations.  The $\pi\pi$ and $\bar KK$ matrix elements at $s=0$ are related to quark-mass derivatives of the corresponding Goldstone boson masses via the Feynman--Hellmann theorem, while the $\pi\eta$ one obeys a Ward identity relating it to a similar vector-current matrix element~\cite{Albaladejo:2015aca}.
At leading order in the chiral expansion we find the normalizations
\be
\Gamma^{I=0}_{\pi\pi}(0)=1,\qquad \Gamma^{I=0}_{KK}(0) = \frac{1}{2}, \qquad \Gamma^{I=1}_{\pi\eta} (0) = \frac{1}{\sqrt{3}}, \qquad  \Gamma^{I=1}_{KK}(0) = \frac{1}{2}. \label{eq:Gamma0leading}
\ee

The next-to-leading order results depend on certain low-energy constants. 
We emphasize that for these, the universality of the relative couplings to different mesons, comparing the scalar form factors and the $S$-waves appearing in the $\Bd$ decays, is not guaranteed, and one might argue in favor of simply using the leading-order relations of eq.~\eqref{eq:Gamma0leading}; see the discussion in appendix~\ref{app:chiral}.  However, in order to obtain at least a realistic estimate of the uncertainties induced by next-to-leading-order corrections, we take those from the scalar form factor matrix elements.
The corresponding low-energy constants are determined in lattice simulations with $N_f=2+1+1$ dynamical flavors at a running scale $\mu = 770\,\MeV$~\cite{Dowdall:2013rya}, limiting the form factor normalizations to the 
ranges\footnote{Note that in ref.~\cite{Daub:2015xja} the form factor normalizations are based on lattice simulations with $N_f=2+1$ dynamical flavors~\cite{Aoki:2013ldr}, which yields similar ranges. In particular, for the fits performed in ref.~\cite{Daub:2015xja} the normalization of the isoscalar kaon form factor was set to $\Gamma^{I=0}_{KK}(0) = 0.6$, which is compatible with the range we use in this analysis.}
\begin{align}
\Gamma^{I=0}_{\pi\pi}(0) &= 0.984 \pm 0.006,& \quad
& \Gamma^{I=1}_{\pi\eta}(0) = (0.56 \dots 0.87), \nonumber \\
\Gamma^{I=0}_{KK}(0) &=(0.44 \dots 0.68),& \quad
&\Gamma^{I=1}_{KK}(0) =(0.38 \dots 0.56). \label{eq:ffnorms}
\end{align}
Since the form factor shape depends on the relative
size of the two pairs of normalization constants, there is some uncertainty in the shape of the isoscalar scalar form factors.
The variations in the isovector form factor normalizations are strongly correlated, i.e.\ their ratio for small and large values of the low-energy constants varies at the 5\% level only.

\section{Coupled-channel and flavor related predictions}\label{sec:resultsB0d}

As explained in the previous section, in the coupled-channel treatment the relative strengths between the $\pi\pi$ or $\pi\eta$ and the isoscalar or isovector $\bar K K$ form factors are fixed, thus the respective final states are related to each other unambiguously. Further, the relative strength between the production amplitudes of different isospin are known.
Hence, since the $\bar B_{d}^0 \to J/\psi \pi^+ \pi^-$ parameters were fitted in ref.~\cite{Daub:2015xja}, we can make predictions for the $\Bd \to J/\psi \pi^0 \eta$ (and the $B^\pm \to J/\psi \pi^\pm \eta$) distribution as well as for the $\bar B_{d}^0 \to J/\psi K^+ K^- / K^0\bar K^0$ $S$-wave amplitudes, which we assume to dominate the differential decay rates in the energy region considered.
The maximal range of this assumed dominance is estimated by predicting the $\pi^0\eta$ $D$-wave as well.

\subsection[$\bar B_{d}^0 \to J/\psi \pi^0\eta $]{\boldmath{$\bar B_{d}^0 \to J/\psi \pi^0\eta$}}\label{sec:pieta}

\begin{figure}
\centering  
\includegraphics*[width=0.49\linewidth]{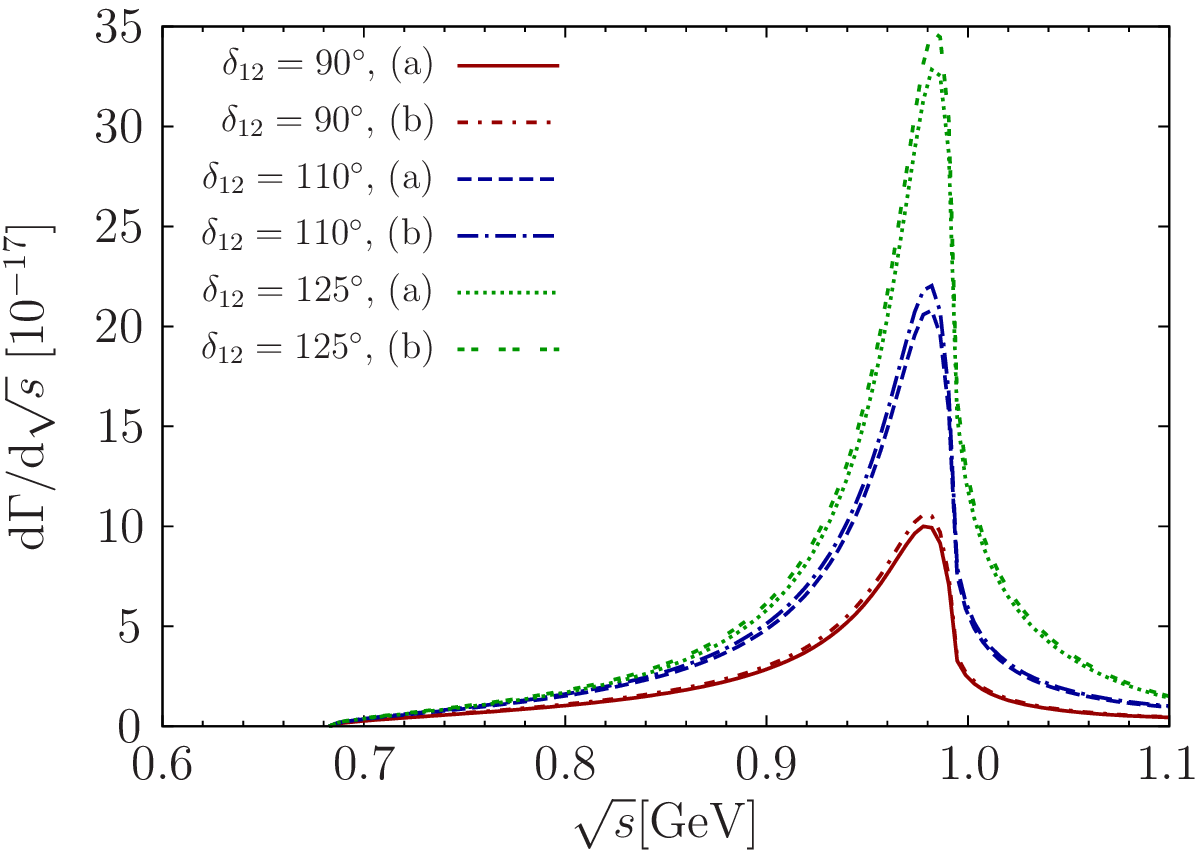} \hfill
\includegraphics*[width=0.49\linewidth]{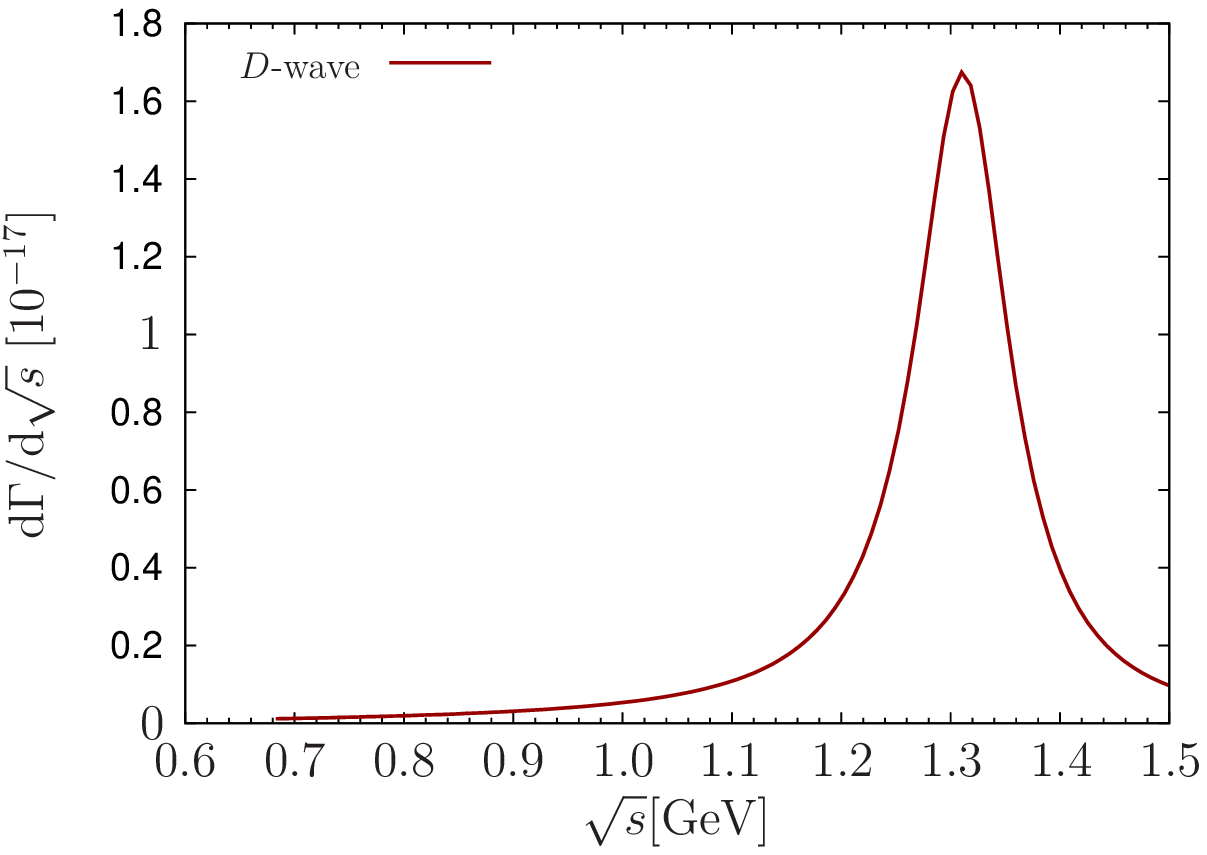}
\caption{Left panel: $\diff\Gamma/\diff\sqrt{s}$ for $\Bd \to J/\psi \pi^0 \eta$. The distribution is plotted for three different input phases $\delta_{12}=90\degree, \, 110\degree,\,125\degree$ and two normalization constants $\bar b_0^n$ due to the fits (a) and (b). The right panel shows the predicted $D$-wave contribution to $\diff\Gamma/\diff\sqrt{s}$ (note the different ranges of the axes).}
\label{fig:gammapieta}
\end{figure}

In figure~\ref{fig:gammapieta} (left panel) the differential decay rate $\diff\Gamma/\diff\sqrt{s}$ for $\Bd \to J/\psi \pi^0 \eta$ is depicted, predicted by means of the fixed isospin relation from the results for two fit configurations (a) and (b) of the $\Bd \to J/\psi \pi^+ \pi^-$ spectrum~\cite{Daub:2015xja}. The distribution is shown for three different input phases $\delta_{12}=90\degree, \, 110\degree,\,125\degree$, limited to an interval compatible with constraints discussed in ref.~\cite{Albaladejo:2015aca}. We see that the dependence on $\delta_{12}$, and hence on details of the $\pi\eta$ interaction, is strong, and completely overwhelms the uncertainty due to the production strength as fixed from the corresponding $\pi^+\pi^-$ channel. 
In order to provide a clearer representation, we refrain from showing uncertainty bands due to the $I=1$ $\pi\eta$ and $\bar K K$ form factor normalizations calculated at next-to-leading order according to eq.~\eqref{eq:ffnorms}. We note that the production strength is fixed from $\Bd \to J/\psi \pi^+ \pi^-$ fits that were performed for those combinations of low-energy constant that result in large isoscalar form factor normalizations~\cite{Daub:2015xja}; we therefore also choose the isovector form factor normalizations in the upper allowed range. This corresponds to similar low-energy constants, though not exactly the same, as we use updated lattice results.  Furthermore, the isovector form factor normalizations depend on additional low-energy constants.
 
Our conclusion is that a measurement of this decay channel will provide important information on the final-state interactions of the $\pi\eta$ $S$-wave system, and in particular the $a_0(980)$ resonance, which dominates this partial wave in the energy range around $1\,\GeV$.

To further substantiate the assumed $S$-wave dominance, we estimate the $\pi\eta$ $D$-wave background, which should become significant in the region of the $a_2(1320)$ resonance.
We model the $a_2$ by a simple Breit--Wigner shape; its coupling strength is related to that of the $f_2(1270)$ by $SU(3)$ symmetry, which decays to $\pi^+\pi^-$ and hence is determined in the $\bar B_{d}^0 \to J/\psi \pi^+ \pi^-$ analysis of ref.~\cite{Aaij:2014siy}. 
In the determination of the $a_2$ strength we employ two ratios: the ratio between the isovector and the isoscalar contributions, eq.~\eqref{eq:dd}, that yields a relative minus sign between the $\bar B_d^0 \to J/\psi f_2$ and the $\bar B_d^0 \to J/\psi a_2$ couplings, as well as the relative strength between the $f_2 \to \pi^+\pi^-$ and the $a_2^0 \to \pi^0 \eta$ couplings. The coupling of a tensor meson to a pseudoscalar pair is obtained from the interaction Lagrangian~\cite{Ecker:2007us, Kubis:2015sga}
\be
\mathcal{L}_{TPP} = g_T \langle T_{\mu\nu} \{u^\mu , u^\nu\} \rangle,
\ee
where $\langle . \rangle$ is the trace in flavor space, $T_{\mu\nu} $ contains the $a_2$ and $f_2$ mesons and $u_\mu = i (u^\dagger \partial_\mu u - u \partial_\mu u^\dagger ) $ the pseudoscalars. As we are interested in the non-strange part of the Lagrangian only, we use for simplicity the $SU(2)$ representations
\begin{equation}
T_{\mu\nu} = \begin{pmatrix}
\frac{a^0_2}{\sqrt{2}}  + \frac{f_2}{\sqrt{6}}  & a^+_2\\
a^-_2& -\frac{a^0_2}{\sqrt{2}} + \frac{f_2}{\sqrt{6}}\\
\end{pmatrix}_{\mu\nu}, 
\quad  
u = \exp{\left(\frac{i \phi }{2F_\pi}\right)}, 
\quad
\phi = \begin{pmatrix}
\pi^0 + \frac{\eta}{\sqrt{3}}  & \sqrt{2} \pi^+\\
\sqrt{2} \pi^-& -\pi^0 + \frac{\eta}{\sqrt{3}}\\
\end{pmatrix},
\end{equation}
where $F_\pi = 92.2\,\MeV$ denotes the pion decay constant. The coupling constant $g_T = 28\,\MeV$ can be obtained consistently from both the $f_2 \to \pi\pi$~\cite{Ecker:2007us} and the $a_2 \to \pi\eta$ decay~\cite{Kubis:2015sga}, confirming $SU(3)$ symmetry.
From the Lagrangian we can finally read off the relative strength of the coupling to $\pi\pi$ and $\pi\eta$ and find $g^2_{a_2 \pi\eta} = g^2_{f_2 \pi\pi}/3$.

The right panel of figure~\ref{fig:gammapieta} shows the predicted $D$-wave. Compared to the $S$-wave contribution shown in the left panel the $D$-wave is negligible in the energy region we consider here, justifying the assumed $S$-wave dominance.

Finally we quote the branching fraction for $\bar B_d^0 \to J/\psi \pi^0\eta$ from the $a_0(980)$ region. 
We integrate the spectrum in the region of the $a_0(980)$ and find
\begin{align}
\B\left(\bar B_{d}^0 \to J/\psi \pi^0\eta\right)\Big|_{\sqrt{s}\,\leq\, 1.1\,\GeV} &= \frac{1}{\Gamma^{\mathrm{tot}}(\bar B_d^0)}\int_{M_\pi +M_\eta}^{1.1\, \GeV} \frac{\diff\Gamma(\bar B_d^0 \to J/\psi \pi^0\eta)}{ \diff\sqrt{s}} \diff\sqrt{s}\nonumber\\
& = \left\{ \begin{array}{rcl} 
(6.0 \dots 6.4) \times 10^{-6} & \quad \mathrm{for}  & \quad \delta_{12} = 90\degree,\\
(1.1 \dots 1.2) \times 10^{-5} & \quad \mathrm{for}   & \quad \delta_{12} = 110\degree, \\
(1.6 \dots 1.7) \times 10^{-5} & \quad \mathrm{for} & \quad \delta_{12} = 125\degree,
\end{array} \right. \label{eq:BRa0}
\end{align}
where the lower and upper values of the given ranges correspond to the $\bar{b}_0^n$ fit results (a) and (b), respectively.
We can compare our results with those of ref.~\cite{Liang:2015qva}. Even if this latter work predicts the $\bar{B}^0_d \to J/\psi \eta \pi^0$ differential decay width without absolute normalization, our distribution can be seen to be narrower. We further note that the numbers in eq.~\eqref{eq:BRa0} are around $3$ to $8$ times larger than the value $(2.2 \pm 0.2) \times 10^{-6}$ estimated in ref.~\cite{Liang:2015qva}, which however only refers to the $a_0(980)$ contribution. To compute it, the authors remove a smooth but large background from the differential decay. Hence, it is quite natural to obtain a larger value for this branching ratio than the one quoted in ref.~\cite{Liang:2015qva}. 

\begin{sloppypar}
The corresponding numbers for $\B(B^\pm \to J/\psi \pi^\pm\eta)\big|_{\sqrt{s}\,\leq\, 1.1\,\GeV}$ are obtained from
eq.~\eqref{eq:BRa0} by multiplying the $\Bd$ branching fractions with a factor of $2.15$, taking into account 
the relative isospin factor of $2$ and a small correction due to the different lifetimes of $B^\pm$ and $\Bd$~\cite{Olive:2016xmw}.
\end{sloppypar}

\subsection[$\bar B_{d}^0 \to J/\psi \bar K K$ $S$-wave prediction]{\boldmath{$\bar B_{d}^0 \to J/\psi \bar  K K$  $S$}-wave predictions}\label{sec:KK}

In this section we discuss our results for the $\bar{B}_d^0 \to  J/\psi \bar{K}K$ decays, where $\bar K K$ can be either a 
neutral or a charged kaon pair. Note that a few data points are available for the latter channel~\cite{Aaij:2013mtm}. 
In ref.~\cite{Aaij:2013zpt} properties of the $f_0(980)$ are deduced based on an amplitude analysis including both resonant and
non-resonant terms performed in the same paper. 
However, on the physical axis a decomposition into resonant and non-resonant contributions is not possible in
a model-independent way. Accordingly the only well-defined quantity that compares the $f_0(980)$ contribution
to the $a_0(980)$ contribution is
\begin{equation}\label{eq:Rf0a0_theo}
\mathcal{R}^{f_0/a_0} = \frac{\int_{4 M_K^2}^{s_p} \big| X^2 \sigma_K \sqrt{s}\, \Gamma_{KK}^{I=0}\big|^2 \diff s}{\int_{4 M_K^2}^{s_p} \big| X^2 \sigma_K \sqrt{s}\, \Gamma_{KK}^{I=1}\big|^2 \diff s} ,
\end{equation}
which is dominated by the two scalar resonances for values of $s_p$ that are not too large.
Our determinations for this quantity are shown in table~\ref{tab:fitYD} for different values of the input phase $\delta_{12}$ and of the upper integration limit $s_p$. 
There is some weak dependence on $s_p$, however, as the table clearly shows, $\mathcal{R}^{f_0/a_0}$ is very sensitive to the input phase $\delta_{12}$.
This sensitivity is even stronger than the uncertainty on the form factor normalizations, which are not accurately determined, cf.\ table~\ref{tab:fitYD}. 
 A measurement of $\mathcal{R}^{f_0/a_0}$ would thus be very valuable to further constrain the so far badly determined  phase~$\delta_{12}$.

\begin{table}[tb]
\noindent
\centering
\begin{tabular}{@{}ccccc@{}}\toprule[1pt]
  $\R^{f_0/a_0}_\text{th}$ & $\delta_{12}= 90\degree $ & $\delta_{12}= 100\degree $ & $\delta_{12}= 110\degree $ & $\delta_{12}= 125\degree $\\
\midrule
$\sqrt{s_p} = 1.05$~GeV & 0.35 \dots 0.46 & 0.25 \dots 0.35 & 0.16 \dots 0.23 & 0.11 \dots 0.15 \\
$\sqrt{s_p} = 1.20$~GeV &  0.33 \dots 0.43 & 0.22 \dots 0.31 & 0.15 \dots 0.22 & 0.11 \dots 0.15\\
\bottomrule
\end{tabular}
\caption{Theoretical determination of the ratio $\R^{f_0/a_0}$, eq.~\eqref{eq:Rf0a0_theo}, for different values of the phase $\delta_{12}$ entering in the determination of the $I=1$ $\bar{K}K$ form factor, $\Gamma_{KK}^{I=1}$, and the upper integration limit $s_p$. The given range for the respective values is due to the uncertainty in the form factor normalizations. 
\label{tab:fitYD}}
\end{table}

\begin{figure}

\includegraphics*[width=0.49\linewidth,keepaspectratio]{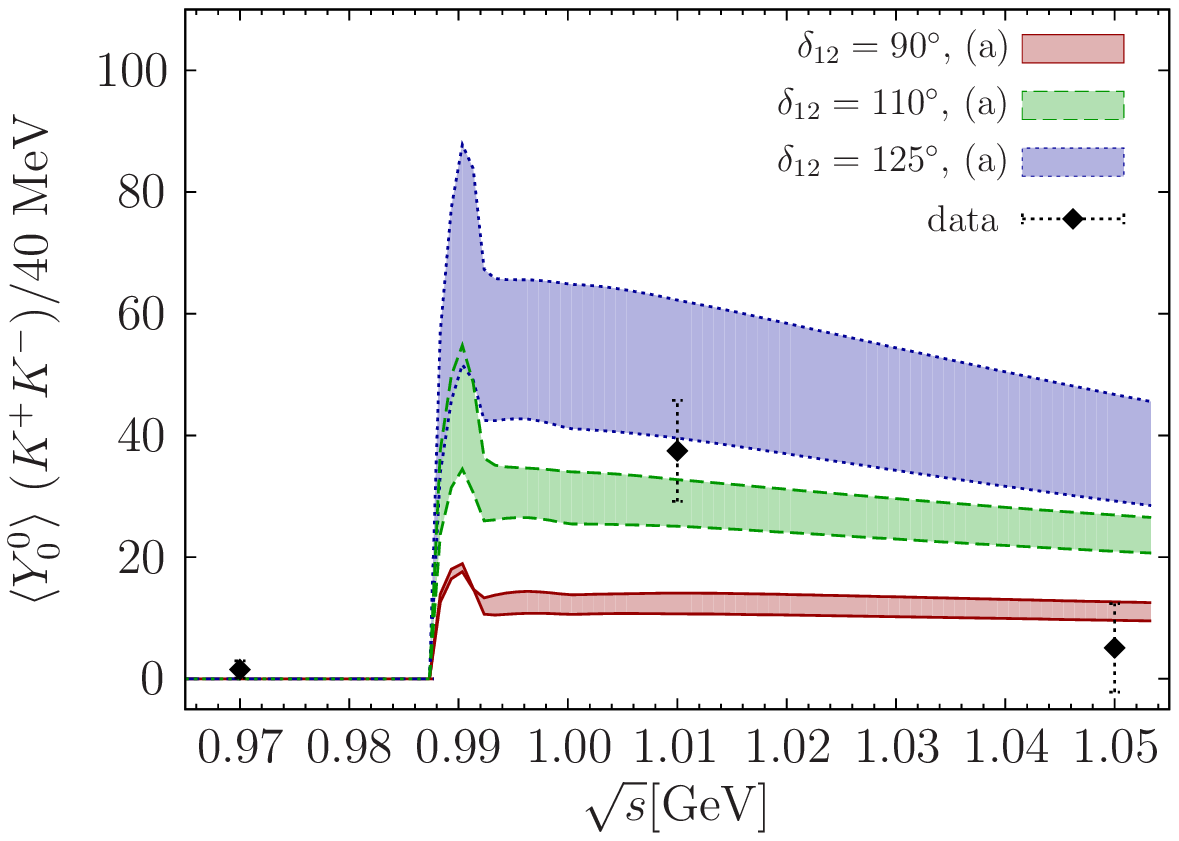}
\hfill
\includegraphics*[width=0.49\linewidth,keepaspectratio]{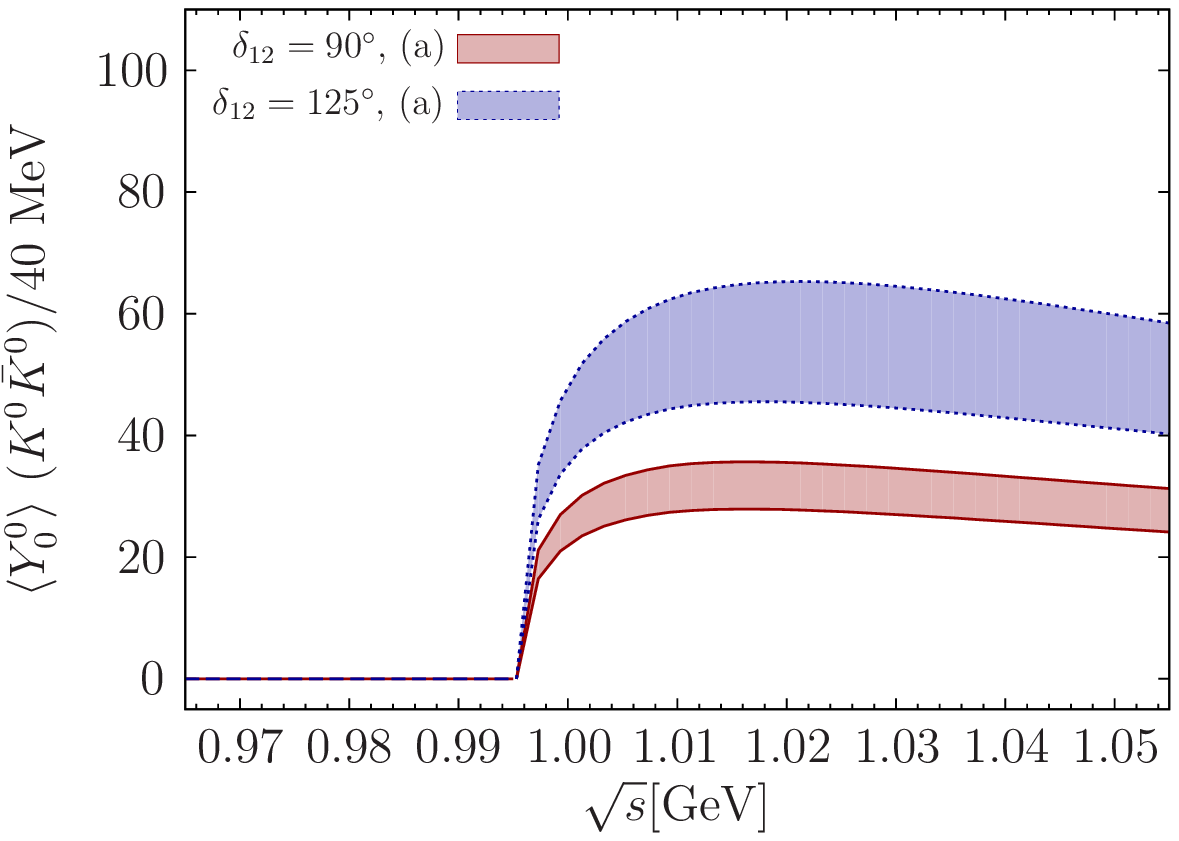}
\\
\includegraphics*[width=0.49\linewidth,keepaspectratio]{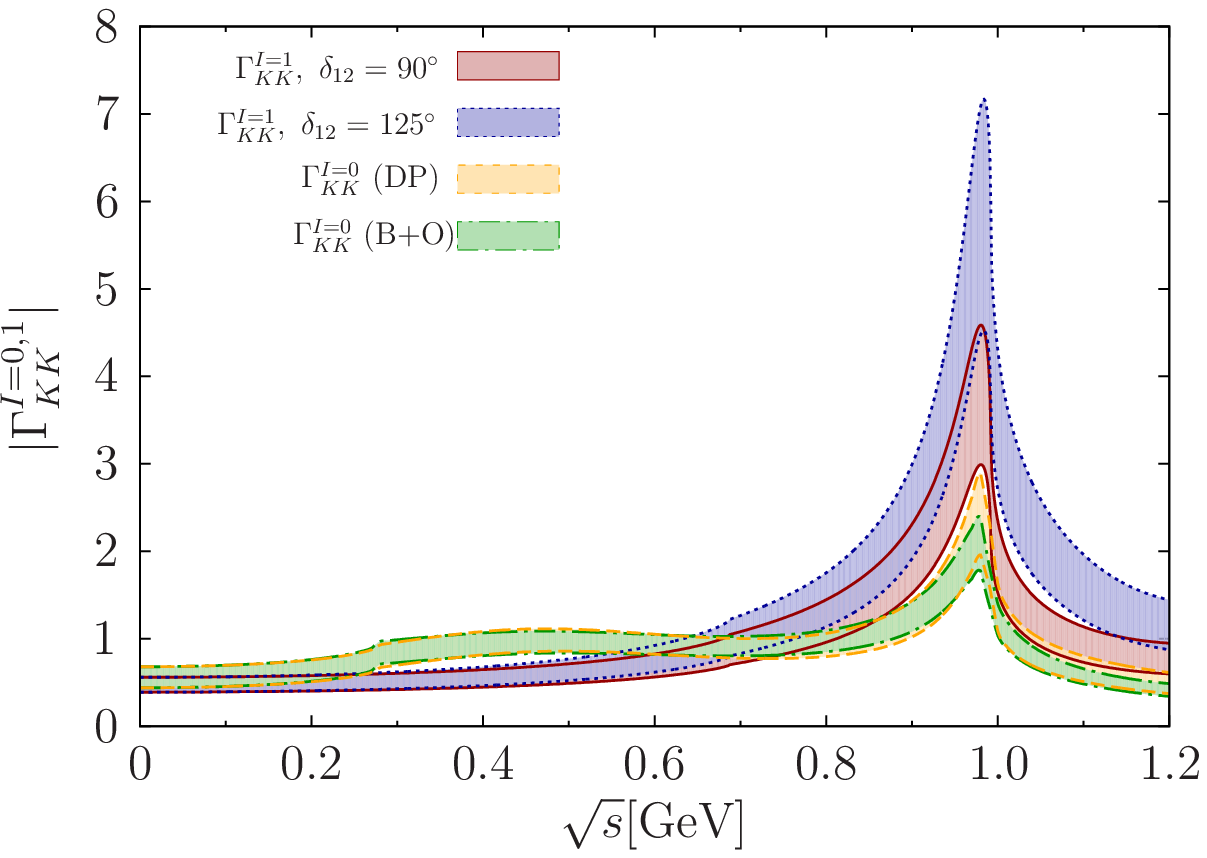}
\hfill
\includegraphics*[width=0.49\linewidth,keepaspectratio]{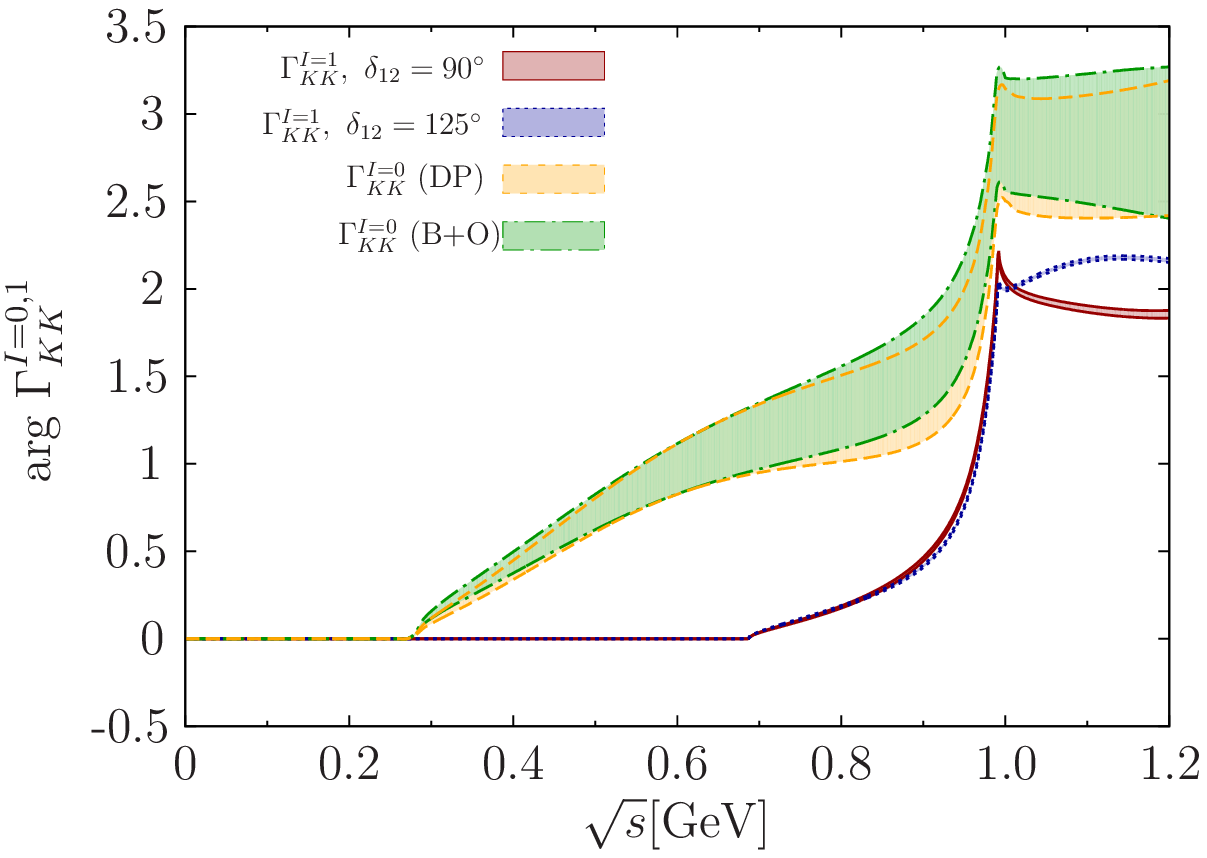}
\caption{Top panels: $\langle Y_0^0 \rangle (s)$ for $\bar{B}^0_d \to J/\psi K^+ K^-$ (left) and $\bar{B}^0_d \to J/\psi K^0 \bar{K}^0$ (right). The curves are calculated for different values of the phase $\delta_{12}$ (see text for further details). The error bands reflect the uncertainties in the normalizations of the form factors. The experimental points are taken from ref.~\cite{Aaij:2013mtm}. Bottom panels: moduli (left) and phases (right) of the scalar form factors $\Gamma_{KK}^{I}$ for $I=0,1$. 
The two bands for the $I=0$ form factor refer to the $T$-matrix solutions of ref.~\cite{Dai:2014zta} (DP) and refs.~\cite{CCL2012,BDM04} (B+O), respectively.
\label{fig:Y00KKbar}}
\end{figure}

Next we turn to our predictions for the physical final states, $K^+K^-$ and $K^0\bar K^0$.\footnote{For simplicity, we discuss the neutral kaon channel in terms of the strong eigenstates.  For even partial waves, the relation to the weak eigenstates, neglecting effects of $CP$-violation, is given by $\diff \Gamma(\bar{B}^0_d \to J/\psi K_S K_S) = \diff \Gamma(\bar{B}^0_d \to J/\psi K_L K_L) = \diff \Gamma(\bar{B}^0_d \to J/\psi K^0 \bar{K}^0)/2$.} In a very naive picture, we would expect a $\bar dd$ 
source term to produce a pair of neutral kaons only, and no charged ones. Equation~\eqref{eq:ffnorms}
in combination with eq.~\eqref{eq:G0s} suggests that at (the unphysical 
point) $s=0$, this naive view is indeed close to reality. If this behavior persisted in the energy region around $1\,\GeV$, we would 
expect a destructive interference of $f_0(980)$ and $a_0(980)$ for charged kaons, but a constructive one for the neutral pairs. 

In figure~\ref{fig:Y00KKbar} we present our results for $\langle Y_{0}^0 \rangle$ for physical $\bar{K}K$ states,\footnote{Note that the decay $\bar{B}^0_d \to J/\psi K^+K^-$ is dominated by the $S$-wave. In the $\bar KK$ $P$-wave there are contributions from the $\rho$ and $\omega$ resonances, which, in principle, can be related to the $\pi\pi$ $P$-wave. These resonances peak below the $\bar KK$ threshold, however, and their contributions are expected to be rather small above. The contribution from the $\phi$ resonance that dominates the $\bar KK$ $P$-wave above threshold is suppressed by the OZI rule, as we are effectively dealing with a $\bar dd$ source.  The suppression of the $P$-wave (and the dominance of the $S$-wave) in this decay has indeed been confirmed experimentally in ref.~\cite{Aaij:2013mtm}. Hence, we here include only the $S$-wave component, and make no attempt to consider $\bar{K}K$ $P$-wave amplitudes.} namely $K^+ K^-$ (left panel) and $K^0 \bar{K}^0$ (right panel), which differ by the sign of the interference term between the isoscalar and the isovector component, according to eq.~\eqref{eq:G0s}.
A simple inspection of the figure shows that the interference pattern anticipated in the previous paragraph does not hold at all. 
To understand the origin of this, in the bottom panel of figure~\ref{fig:Y00KKbar} we show the two ($I=0,1$) kaon form factors: we see that the $I=1$ one completely dominates over the $I=0$ form factor, or in other words, the $a_0(980)$ signal is much stronger than the $f_0(980)$ one; hence, both the $K^+K^-$ and $K^0\bar{K}^0$ angular moments are dominated by the $I=1$ form factor, which explains why the interference pattern is not the one naively expected. 
The small value theoretically obtained for $\R^{f_0/a_0}$ is a direct consequence of the $I=1$ dominance found here, as can be seen from its definition, eq.~\eqref{eq:Rf0a0_theo}. 
The difference in strengths of the two resonances is somewhat remarkable.  
It is \emph{not} due to significant differences in the pole positions, as the $I=1$ $T$-matrix of ref.~\cite{Albaladejo:2015aca}
uses an $a_0(980)$ pole position (on the second sheet) of 
\begin{equation}\label{eq:a0pole}
\sqrt{s_{a_0(980)}^{\mathrm{II}}} = \big(994 \pm 2 - i (25.4 \pm 5.0)\big) \,\MeV
\end{equation}
as a constraint~\cite{Isidori:2006we,Ambrosino:2009py}, while the second-sheet pole of the $f_0(980)$ corresponding to our (main) 
$I=0$ $T$-matrix parametrization is given by~\cite{Moussallam:2011zg}
\begin{equation}\label{eq:f0pole}
\sqrt{s_{f_0(980)}^{\mathrm{II}}} = \Big(996^{+4}_{-14}  - i \big(24^{+11}_{-3}\big)\Big) \,\MeV ,
\end{equation}
determined from pion--pion Roy equations.  It seems, therefore, that rather the residues of the resonance couplings to the respective 
currents are very different, which might be in conflict with at least a simple interpretation of both scalar resonances
as $\bar KK$ molecules: in such a picture, similar binding energies would imply similar coupling strengths
to the $\bar KK$ channel according to the Weinberg criterion~\cite{Weinberg:1962hj,Baru:2003qq}, at least as long as the coupling
to the second channel ($\pi\pi$ and $\pi\eta$) is weak enough to be perturbative.\footnote{Note that the central values for the
pole positions quoted in eqs.~\eqref{eq:a0pole}, \eqref{eq:f0pole} also slightly discourage such a simple picture, as they lie 
somewhat \emph{above} the $\bar KK$ threshold.}  More detailed investigations of the couplings of the scalars to $\bar qq$ operators,
completing ref.~\cite{Moussallam:2011zg} also in the $I=1$ sector, would be very interesting to clarify this issue.

It is remarkable to note that the \textit{phase} of the two form factors, see bottom right panel of figure~\ref{fig:Y00KKbar},
changes almost perfectly in parallel, starting from the $\pi\eta$ threshold---the only difference is the low-energy phase growth
in the isospin-0 channel, associated with the $f_0(500)$.  In a single-channel phase-dispersive \Omnes representation, 
the isospin-0 will then roughly equal the isospin-1 form factor, multiplied with a second \Omnes function that contains the 
low-energy phase rise between $\pi\pi$ and $\pi\eta$ thresholds only. 
It is easily seen that this factor, leading to the 
famous $f_0(500)$ enhancement, simultaneously depletes the resonance signal of the $f_0(980)$ quite significantly:
while it is (broadly) peaked around $\sqrt{s}=500\,\MeV$, it falls well below $1$ around $1\,\GeV$, the region of the 
second resonance. 

Finally, it can be seen in figure~\ref{fig:Y00KKbar} that the peak in the $K^+K^-$ angular moment is located around $\sqrt{s} \simeq 990\,\MeV$ and is quite narrow, whereas the $K^0\bar{K}^0$ threshold lies at $\sqrt{s} \simeq 995\,\MeV$. 
Accordingly the peak in the $K^0\bar{K}^0$ distribution is phase-space suppressed.

There are three experimental points available in the charged kaon spectrum in the region we investigate, providing a test of our prediction, which is solely based on coupled-channel and isospin relations.\footnote{As discussed in the previous section the experimental spectra of the decays with $K^+ K^-$ and $\pi^+\pi^-$ meson pairs are not equally normalized. Thus once the predicted $K^+K^-$ angular moment, whose calculation is based on fit results of the $\pi^+\pi^-$ spectrum, is compared directly to data, an appropriate normalization between the $K^+ K^-$ and $\pi^+\pi^-$ spectra has to be achieved, in addition to all kinematical replacements. For details see ref.~\cite{Daub:2015xja}. Note also that the data we compare to is binned; to compare the spectrum with charged to the one with neutral kaon pairs in the final states we rescale the latter.}
Each of the two data points that are located above threshold agrees with one of our different determinations of $\langle Y_0^0 \rangle$, depending on the value of the input phase $\delta_{12}$. Tempting as it would be, we refrain from fitting these data points by modifying our theoretical input. We content ourselves with illustrating that new and improved data would help in constraining the $\pi\eta$--$\bar{K}K$ form factors, and thus our theoretical understanding of their final-state interactions and the properties of the $a_0(980)$ and (indirectly) the $f_0(980)$ resonances. 

\section{Summary and outlook}\label{sec:summary}

In this paper we demonstrated that experimental data for $\bar{B}^0_d \to J/\psi(\pi\eta,\bar{K}K)$ can be used to further constrain a parameter, $\delta_{12}$,
that is crucial to pin down the $\pi\eta$ scattering amplitude. Around $1\,\GeV$, the latter is dominated by the pole of the scalar meson $a_0(980)$.
A high-accuracy determination of $\delta_{12}$ would, however, need further theoretical development. 

In addition, in the scalar sector enhanced isospin-violating effects can occur around the two-kaon thresholds, driven by both the proximity of
resonances in the isoscalar ($f_0(980)$) as well as isovector ($a_0(980)$) channels, and the $8\,\MeV$ gap between the $K^+K^-$
and $K^0\bar K^0$ thresholds~\cite{Achasov:1979xc}. This phenomenon  is usually referred to as $a_0$--$f_0$ mixing in the literature,
and has been argued to be significant for, e.g., $\eta(1405)\to 3\pi$~\cite{Wu:2011yx,Aceti:2012dj},  weak decays of $D_s/B_s$ mesons~\cite{Wang:2016wpc},
and $J/\psi\to \phi \pi^0\eta$~\cite{Wu:2007jh,a0f0}.
The predictions of the last mentioned theoretical calculations were confirmed experimentally at BES-III~\cite{Ablikim:2010aa}.

First steps towards a rigorous dispersive treatment of $a_0$--$f_0$ mixing are reported in ref.~\cite{Albaladejo:2015bca,Albaladejo:2017hhj}. An adaption of this formalism to
the reactions at hand will be pursued elsewhere. Here it would in particular be important to perform a detailed study of
$\bar B^0_s\to J/\psi \pi^0\eta$, since the weak decay that drives the transition leads to a purely isoscalar source.

\acknowledgments
We are grateful to Michael Pennington for providing us with the coupled-channel $T$-matrix parametrization
of ref.~\cite{Dai:2014zta}.
Financial support by DFG and NSFC through funds provided to the Sino--German CRC~110
``Symmetries and the Emergence of Structure in QCD''
is gratefully acknowledged.
M.~A.\ acknowledges financial support from the Spanish MINECO and European FEDER funds under the contracts 27-13-463B-731 (``Juan de la Cierva'' program), FIS2014-51948-C2-1-P, FIS2014-57026-REDT, and SEV-2014-0398, and from Generalitat Valenciana under contract PROMETEOII/2014/0068.

\appendix
\section{Flavor relations from a chiral Lagrangian for \boldmath{$B_d^0\to J/\psi M_1 M_2$}}\label{app:chiral}

We consider the decays $B_d^0\to J/\psi M_1 M_2$, where $M_i$ is a light
pseudoscalar meson ($\pi$, $K$, $\eta$).  We assume that
there exists a kinematical regime where the light mesons are soft, and that we
can describe the dynamics in this situation via a chiral expansion.  We can
write a chiral Lagrangian that reflects the chiral transformation properties
of the weak transition operator following the method 
described e.g.\ in ref.~\cite{Bijnens:2009yr} for $K\to \pi\pi$ and $K\to 3\pi$ decays,
where the $K$ was assumed to be heavy. The two dominant operators have the following
structure:
\begin{equation}
O_d\sim \bar{c}_L \gamma_\mu d_L \, \bar{b}_L \gamma^\mu c_L    ,
\end{equation}
such that, under the chiral symmetry group $SU(3)_L\times SU(3)_R$, they
transform simply as $d_L$. In order to construct a chiral Lagrangian we introduce a
three-vector spurion field $t_L$, which transforms as 
\begin{equation}
t_L \to g_L\, t_L , \qquad g_L \in SU(3)_L ,
\end{equation}
such that $t_L^\dagger (O_u,O_d,O_s)$ is a chiral invariant.
$t_L$ will ultimately be set to $t_L=(0,1,0)^t$. The heavy vector field
$\Psi^\mu$ is left invariant by chiral symmetry, while the  $B_d^0$ can be
considered as part of a three-vector $B = (B^+,B_d^0,B_s^0)^t$,
which transforms as
\begin{equation}
B\to h\,B ,
\end{equation}
where $h$ is the non-linear realization of the chiral group. That is,
if $U$ is the chiral field matrix and $U=u^2$,
\begin{equation}
U\to g_R\, U\, g_L^\dagger,\qquad
u \to g_R\, u\, h^\dagger = h\, u\, g_L^\dagger , \qquad
g_R \in SU(3)_R .
\end{equation}
At leading chiral order the Lagrangian that describes the decays 
$B_d^0 \to J/\psi \,+\,\hbox{light mesons}$ has two independent terms:
\begin{equation}\label{eq:Lag_1}
{\cal L}_1 = g_{1a}\, t^\dagger_L\, u^\dagger u_\mu  B\,\Psi^\mu
+ i g_{1b}\, t^\dagger_L\,u^\dagger \nabla_\mu  B\,\Psi^\mu + \text{h.c.} ,
\end{equation}
with
\begin{equation}
u_\mu=i(u^\dagger\partial_\mu{u} - u\partial_\mu{u^\dagger}), \qquad
\nabla_\mu\, B=(\partial_\mu+\Gamma_\mu) B, \qquad
\Gamma_\mu =\frac{1}{2}(u^\dagger\partial_\mu{u} + u\partial_\mu{u^\dagger}) ,
\end{equation}
using the same notation as e.g.\ in ref.~\cite{Ecker:1988te}.
We are interested in the production of meson pairs: expanding the
Lagrangian~\eqref{eq:Lag_1} to quadratic order in the light fields and using integration by parts, we find
\begin{align}\label{eq:Lagphi^2}
{\cal L}_{1,B_d^0}^{\phi^2}  = - \frac{ig_{1}}{4F_\pi^2} \Psi^\mu
\bigg\{& \partial_\mu B_d^0
\Big[
\frac{1}{2}(\pi^0)^2 + \pi^+ \pi^- 
-\frac{1}{\sqrt3} \pi^0\eta
+\frac{1}{6}\eta^2
+ \Kz\Kzb 
\Big] \nonumber \\
& + 
B_d^0 \Big[ \pip \partial_\mu\pim - \pim\partial_\mu\pip
+ \Kzb \partial_\mu\Kz - \Kz\partial_\mu\Kzb
\Big] \bigg\}+ \text{h.c.}  ,
\end{align}
with
$g_1=2g_{1a}+g_{1b}$.
The second line in eq.~\eqref{eq:Lagphi^2} contributes to amplitudes where the
$M_1M_2$ pair is in a $P$-wave (note that there is no contribution to
$\pi\eta$ at this order). The first line in eq.~\eqref{eq:Lagphi^2} contributes to
amplitudes where the pair is in a relative $S$-wave.  This gives a set of
definite relations among $\pi\pi$, $\pi\eta$, and $\Kbar K$ $S$-wave
amplitudes. These are exactly the same as for the $\bar{d}d$ scalar form
factors at chiral order $p^2$. Indeed, the $\bar{d}d$ form factors at leading
order are obtained by expanding the $\mathcal{O}(p^2)$ chiral Lagrangian piece
\begin{equation}
{\cal L}_2= \frac{F_\pi^2}{4}\,\braque{\chi_+} ,
\end{equation}
with $\chi_+ = u^\dagger\chi u^\dagger + u \chi^\dagger u$, $\chi=2\B_0\,\text{diag}(0,1,0)$. 
Similar arguments to derive such relations have also been formulated in ref.~\cite{Oset:2016lyh}.

In addition we can write the contributions to the decays of the $B^+$ meson, 
$B^+\to J/\psi M_1 M_2$, as
\begin{align}\label{eq:LagphiBd}
{\cal L}_{1,B^+}^{\phi^2}  = - \frac{ig_{1}}{4F_\pi^2} \Psi^\mu &\bigg\{
\partial_\mu B^+ \bigg[
\sqrt{\frac{2}{3}}\pim\eta + K^0 K^- \bigg] \nonumber \\
& +  B^+ \Big[
 \sqrt2 (\piz\partial_\mu\pim-\pim\partial_\mu\piz)
+ \Km\partial_\mu\Kz - \Kz\partial_\mu\Km \Big]
\bigg\}+ \text{h.c.} .
\end{align}

The above derivation is rather general: it does not rely on
factorization, the large-$N_c$ expansion, or other hypotheses. 
The relations obtained are, however, valid only at leading
order. Indeed, at next-to-leading order, the one-loop divergences of the $\bar{d}d$ form
factors are absorbed into the standard chiral coupling
constants $L_i$~\cite{Gasser:1984gg}, while the one-loop divergences of the $B_d^0$ amplitudes are
absorbed into coupling constants $g_{3a}$, $g_{3b}$, \ldots, pertaining to the
higher-order generalization of the Lagrangian~\eqref{eq:Lag_1}.  These are
obviously unrelated to the couplings $L_i$. It is also likely that the chiral
logarithms of the $B_d^0$ amplitudes will be different from those of the $\bar{d}d$
form factors.

\section{Estimates for the \boldmath{$\pi\eta$ $P$-waves} and left-hand cuts}
\label{app:Pwaves}

\subsection{Chiral Lagrangians}\label{app:Pwave-chiral}

In order to generate a non-vanishing $\pi\eta$ $P$-wave contribution, 
the chiral Lagrangian needs to involve an explicit symmetry-breaking mass term 
$\propto \chi_\pm = u^\dagger\chi u^\dagger \pm u \chi^\dagger u$, where
$\chi=2\B_0\,\text{diag}(m_u,m_d,m_s)$ is proportional to the quark mass matrix.  
The lowest-order chiral Lagrangian that can produce a $\pi\eta$ pair in a $P$-wave is
\begin{equation}
{\cal L}_3= \frac{g_3}{8} t_L^\dagger u^\dagger B\, \psi^\mu 
\braque{\chi_+\, u_\mu} + \frac{g'_3}{16} t_L^\dagger u^\dagger u_\mu B\, \psi^\mu 
\braque{\chi_-} + \text{h.c.}  . \label{eq:Lagr3}
\end{equation}
An expansion of ${\cal L}_3$ to quadratic order in the light fields yields
\begin{equation}
{\cal L}^{\phi^2}_3 =  \frac{i(M_K^2-M_\pi^2)}{4\sqrt3 F_\pi^2}
 \left[   ( g_3 + g'_3) \partial_\mu( \pi^0 \eta)
+ (- g_3 + g'_3) ( \partial_\mu\pi^0 \eta - \partial_\mu\eta \pi^0)
\right]
 B_d\, \psi^\mu + \ldots + \text{h.c.}, 
\end{equation}
where the ellipsis denotes terms involving other meson pairs than $\pi\eta$. 
The last term contributes to a $P$-wave amplitude, which we can compare to the $S$-wave:
\begin{align}
{\cal M}_{3,P}&= \frac{(g_3-g'_3)(M_K^2-M_\pi^2)}{4\sqrt3F_\pi^2}
\,(p_1^\mu -p_2^\mu)\epsilon_{\mu}^*\,,\nonumber\\
{\cal M}_{1,S}&= \frac{g_1 - (g_3+g'_3)(M_K^2-M_\pi^2)}{4\sqrt3 F_\pi^2} 
\,(p_1^\mu +p_2^\mu)\epsilon_{\mu}^*\,. \label{eq:M1S+M3P}
\end{align}
We express the matrix elements in the basis of the momentum vectors eq.~\eqref{eq:newbasis}, using
\begin{align}
P^\mu &= (p_1^\mu + p_2^\mu) = \frac{P\cdot p_\psi}{m_\psi^2}\, p^\mu_\psi + P^\mu_{(0)} , \nonumber\\
Q^\mu &=(p_1^\mu - p_2^\mu)=\frac{Q\cdot p_\psi}{m_\psi^2}\, p^\mu_\psi 
+\left(\frac{Y\left(P \cdot p_\psi\right)}{X} \cos\theta_\eta -\frac{M_\eta^2-M_\pi^2 }{s}\right) P^\mu_{(0)}
+ Q^\mu_{\parallel} . \label{eq:PQrep}
\end{align}
A natural order of magnitude estimate for the ratio of the chiral
coupling constants is
\begin{equation}
\frac{g_3-g'_3}{g_1}\sim \frac {1}{\Lambda^2},\qquad \Lambda\simeq 4\pi F_\pi \simeq 1\,\GeV . 
\end{equation}
Putting pieces together, we estimate the $P$-wave-to-$S$-wave ratio in the amplitude ${\cal F}_0$ 
\begin{equation}
\frac{{\cal F}_0^{(P)}(s)}{{\cal F}_0^{(S)}(s)}=
\frac{ (M_K^2-M_\pi^2)\, Y (P \cdot p_\psi)}{\sqrt{3}\,\Lambda^2 X} + \ldots,
\end{equation}
and replacing the kinematic functions $X$, $Y$, and $P \cdot p_\psi = (m_B^2-s-m_\psi^2)/2$, we arrive at
\begin{equation}\label{chiralS/P}
\frac{{\cal F}_0^{(P)}(s)}{{\cal F}_0^{(S)}(s)}=
\frac{ (M_K^2-M_\pi^2)}{\sqrt{3}\,\Lambda^2}
\frac{\lambda^{1/2}(s,M_\eta^2,M_\pi^2)}{s} + \ldots,
\end{equation}
where we have neglected terms of higher order in the chiral expansion.

While this is derived from a chiral Lagrangian, it is plausible that
the chiral estimate for the $P$-wave should be valid up to $\sqrt{s} \simeq 1\,\GeV$ due to the absence of final-state interactions.  
Since the
final-state interactions for the $S$-wave increase its value significantly, we can
derive an upper bound for the ratio from~\eqref{chiralS/P}: 
\begin{equation}
\vert {\cal F}_0^{(P)}(s)/{\cal F}_0^{(S)}(s)\vert \lesssim 0.05 ,
\end{equation}
which should be valid in the region $ \sqrt{s} \le 1\,\GeV$.

It is obvious from eqs.~\eqref{eq:M1S+M3P} and \eqref{eq:PQrep} that the Lagrangian $\mathcal{L}_3$ also produces
a $P$-wave in the transversity form factor $\F_\parallel$.  Just for completeness, we in addition show a Lagrangian term
that generates a $P$-wave in the remaining form factor $\F_\perp$:
\begin{equation}
\mathcal{L}_{4\perp} = \frac{g_{4\perp}}{8} t_L^\dagger \, u^\dagger \, \epsilon^{\mu\nu\alpha\beta} u_\mu  \nabla_\alpha B \, \psi_\beta 
\braque{\chi_+ u_\nu} .
\end{equation}
This also involves an explicit symmetry-breaking mass term, however, 
as indicated by the notation, it is of higher chiral order than the terms in eq.~\eqref{eq:Lagr3}.

\subsection[$\psi(2S)$-exchange]{\boldmath{$\psi(2S)$-exchange}}\label{app:psi2S}

\begin{figure}
\centering  
\includegraphics[scale=0.6]{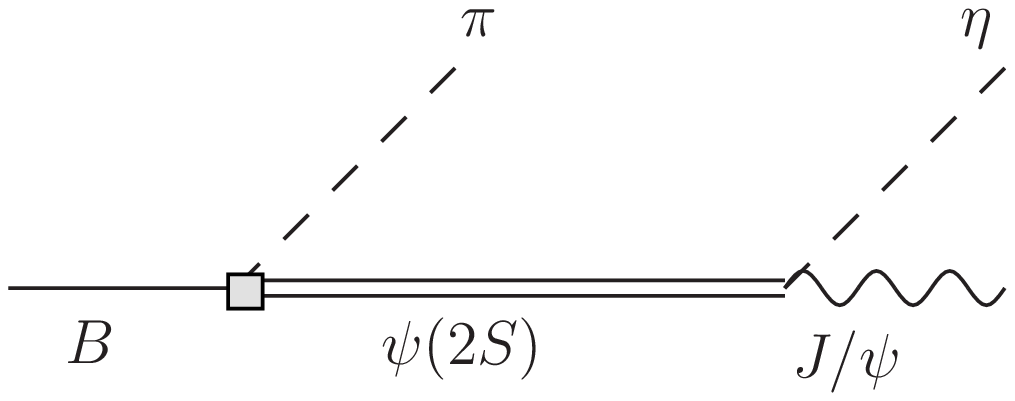}
\hfill
\includegraphics[scale=0.6]{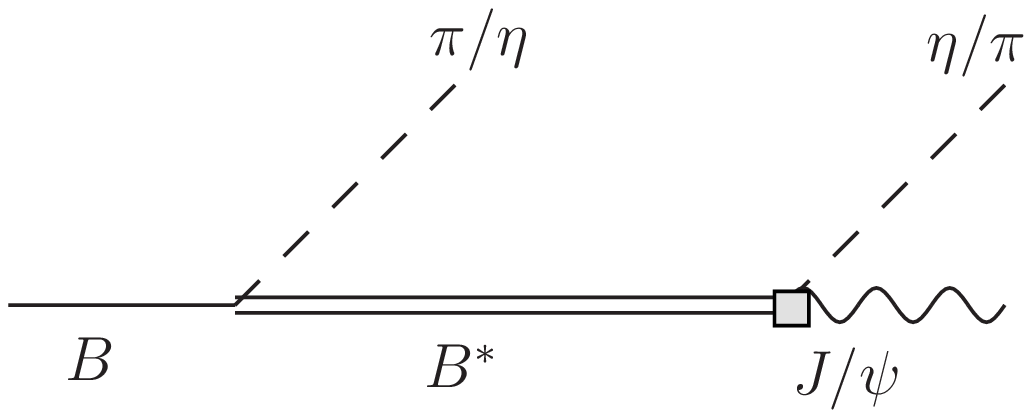}
\caption{Left panel: $t$-channel $\psi(2S)$ exchange diagram; right panel: $t/u$-channel diagrams for a $B^*$ exchange.
The weak decay vertex is marked by a gray square in both cases.
Note that $u$-channel $\psi(2S)$ exchange is negligible due to the isospin suppression of the decay $\psi(2S) \to J/\psi \pi^0$.}
\label{fig:exchange}
\end{figure}

In this appendix, we calculate the contribution of $t$-channel exchange of the $\psi(2S)\equiv\psi'$ resonance to the 
decay amplitude $\Bd(p_B)\to J/\psi(p_\psi)\pi^0(p_1)\eta(p_2)$ as depicted in figure~\ref{fig:exchange} (left panel).  
We write the vertex for $\Bd(p_B)\to\psi'(q)\pi^0(p_1)$ in terms of an effective coupling constant $\bgB$ as
\be
\frac{G_F}{\sqrt{2}}V_{cb}V^*_{cd} f_\psi m_\psi \,\bgB\, (p_B+p_1)_\mu \epsilon^{\mu *}(q,\nu) ,
\ee
where $\epsilon_\mu^*(q,\nu)$ denotes the polarization vector of the $\psi'$ with helicity $\nu$,
which yields a partial width
\be
\Gamma(\Bd \to \psi' \pi^0) = \frac{G_F^2 |V_{cb}|^2|V_{cd}|^2f_\psi^2m_\psi^2}{2}\frac{|\bgB|^2}{16\pi} \frac{\lambda^{3/2}\big(m_B^2,\mpsi,M_\pi^2\big)}{m_B^3 \mpsi} .
\ee
From the branching fraction $\B\big(\Bd\to\psi'\pi^0\big) = (1.17\pm0.19)\times 10^{-5}$~\cite{Chobanova:2015ssy} and the life time $\tau_{\Bd} = (1.519\pm0.005)\times 10^{-12}\,\text{s}$~\cite{Olive:2016xmw}, we find
\be
|\bgB| \approx 0.14.
\ee
The subsequent decay $\psi'(q)\to J/\psi(p_\psi)\eta(p_2)$ is parametrized in terms of an amplitude
\be
i\,\gpsi \epsilon_{\mu\nu\alpha\beta}\, \epsilon^{\mu *}(p_\psi,\lambda)\,p_\psi^\nu \,\epsilon^\alpha(q,\nu)\,q^\beta ,
\ee
leading to a partial width
\be
\Gamma\big(\psi'\to J/\psi\eta) = \frac{\gpsi^2}{96\pi} \frac{\lambda^{3/2}\big(\mpsi,m_\psi^2,M_\eta^2\big)}{m_{\psi'}^3} .
\ee
From the branching fraction $\B(\psi'\to J/\psi\eta) = (3.36\pm0.05)\%$ and the total width 
$\Gamma(\psi') = (296\pm 8)\,\keV$~\cite{Olive:2016xmw}, we deduce the coupling $|\gpsi| \approx 0.218\,\GeV^{-1}$.
Altogether, $\psi'$-exchange leads to a contribution to the $\Bd\to J/\psi \pi^0\eta$ transversity form factor
$\F_\perp^{(\psi')}$ of the form 
\be
\F_\perp^{(\psi')}(s,t) = \frac{\bgB\gpsi \sqrt{s}X}{t-\mpsi} .
\ee
The leading $P$-wave can be obtained from the partial-wave expansion~\eqref{eq:F-PWE}, we find
\be
\F_\perp^{(\psi')(P)}(s) = \sqrt{\frac{3}{2}} \bgB\gpsi \frac{\sqrt{s}}{Y} \bigg( w - \frac{w^2-1}{2}\log\frac{w+1}{w-1} \bigg), \label{eq:FperpP}
\ee
with the kinematical variables $X$, $Y$ as defined in the main text, and 
\be
w = \frac{1}{2XY}\bigg[ \Sigma -2\mpsi -s+\frac{\Delta_m}{s} \bigg] , 
\quad \Sigma = m_B^2+m_\psi^2+M_\pi^2+M_\eta^2 .
\ee
Note that $w>1$ \textit{except} for the interval $s \in [ s_1,s_2]$,
\begin{align}
s_{1/2} &= \frac{1}{2}\Bigg[ \Sigma-\mpsi-\frac{(m_B^2-M_\pi^2)(m_\psi^2-M_\eta^2)
\pm \lambda^{1/2}\big(m_B^2,\mpsi,M_\pi^2\big)\lambda^{1/2}\big(\mpsi,m_\psi^2,M_\eta^2\big)}{\mpsi}\Bigg] 
\nonumber\\
&\approx \big\{(1.34\,\GeV)^2 , (1.83\,\GeV)^2\big\} .
\end{align}
The $\psi(2S)$-exchange mechanism contributes by far most of its strength in this interval, where 
the resonance can go on-shell, and the partial-wave approximation~\eqref{eq:FperpP} is insufficient; 
however, the integrated partial width due to this mechanism, see eq.~\eqref{eq:psi2SBR}, 
is two orders of magnitude smaller compared to the one integrated in the $a_0(980)$ region, eq.~\eqref{eq:BRa0}.
In addition, at energies around $1\,\GeV$, in contrast, the $P$-wave fully dominates the $\psi(2S)$ exchange,
and contributes to the differential decay rate according to
\begin{equation}
\frac{\diff\Gamma}{\diff\sqrt{s}} \bigg|_{\psi(2S)} =  \frac{G_F^2 |V_{cb}|^2 |V_{cd}|^2 f_\psi^2 m_\psi^2 X Y \sqrt{s}}{2 (4\pi)^3 m_B^3}  \Big|Y^2 \F_\perp^{(\psi')(P)}(s) \Big|^2
.
\end{equation}
We find this to be smaller than the $a_0(980)$ signal by about five orders of magnitude:
this particular contribution to the $P$-wave as well as to the left-hand cut of the process 
is entirely negligible.

\subsection[$B^*$-exchange]{\boldmath{$B^*$}-exchange}\label{app:B*}

An alternative mechanism generating a left-hand-cut structure is given by the exchange of a $B^*$ meson in 
either the $t$- or $u$-channel, see figure~\ref{fig:exchange} (right panel).  In contrast to the $\psi(2S)$-exchange discussed in the previous section,
the $B^*$ cannot go on-shell in the decay, therefore the associated left-hand cut is outside the physical
decay region.  On the other hand, the exchange of a $B^*$ is not suppressed in any obvious manner 
(such as by the OZI mechanism), hence it is potentially much more sizable.
The coupling of a $B/B^*$ to a light pseudoscalar is given by the Lagrangian term~\cite{Wise:1992hn}
\begin{equation}
\frac{g}{2} \text{Tr}\big[ \bar H_a H_b \gamma_\nu \gamma_5 \big] u^\nu_{ba} , \label{eq:BBpi}
\end{equation}
where $\text{Tr}[\ldots]$ denotes the Dirac trace, $a,\,b$ are flavor indices, 
and $H = \frac{1}{2}(1+\slashed{v})[B^*_\mu \gamma^\mu - B \gamma_5]$ is the covariant field combining the pseudoscalar and vector $B$ mesons $(B^{-(*)},\bar B^{0(*)}, \bar B_s^{(*)})$ of velocity $v$, taken to be mass-degenerate in the heavy-quark limit. 
These fields are of mass dimension $3/2$ as factors of $\sqrt{m_B}$ and $\sqrt{m_{B^*}}$ are absorbed in the $B^*_\mu$ and $B$ fields.
Heavy-flavor symmetry dictates that the same coupling $g$ also determines the couplings of charmed $D/D^*$ mesons
to light pseudoscalars [with $m_B\to m_D$ in~\eqref{eq:BBpi}]; the resulting partial width
\begin{equation}
\Gamma\big(D^{+*}\to D^+\pi^0\big) = \frac{g^2 m_D^2}{192\pi F_\pi^2}\frac{\lambda^{3/2}(m_{D^*}^2,m_D^2,M_\pi^2)}{m_{D^*}^5}
\end{equation}
allows one to pin down the coupling $g \approx 0.58$.  
For the weak vertex $B^* \to J/\psi M_2$, $M_2=\pi^0, \, \eta$, we use a Lagrangian with four different trace structures obeying the desired transformation behavior under heavy-quark spin symmetry~\cite{Mannel:1990un},
\begin{align}
& \tilde G_F \,  t_L^\dagger u^\dagger \Big\{ 
\text{Tr}\left[ (\alpha_1 + \alpha_2 \slash{v'})
\gamma^\mu(1-\gamma^5) J
\gamma_\mu(1-\gamma^5) \bar{H} \right]\nonumber\\
& \qquad\quad +  \hbox{Tr}\left[ 
\gamma^\mu(1-\gamma^5) J\right]\text{Tr}\left[(\beta_1 + \beta_2\slash{v'})
\gamma_\mu(1-\gamma^5) \bar{H} \right] \Big\}, \nonumber\\ 
& \tilde G_F = \frac{G_F}{\sqrt{2}}V_{cb}V^*_{cd} f_\psi m_\psi .
\end{align}
$J = \frac{1}{2}(1+\slashed{v'})[\Psi_\mu \gamma^\mu - \eta_c \gamma_5]$ 
combines the lightest pseudoscalar ($\eta_c$) and vector ($J/\psi$) charmonium fields that carry velocity $v'$. Similarly to the $B$-meson fields they are of mass dimension 3/2 and taken to be mass-degenerate in the heavy-quark limit. 
We therefore can relate the required $B^* \to J/\psi M_2$ vertex to decays $B^0 \to J/\psi M_2$ and $B^0 \to \eta_c M_2$,
\begin{align}
\Gamma\big(B^0 \to J/\psi \pi^0\big) &= \frac{\tilde G_F^2}{4 F_\pi^2} 
\frac{\lambda^{1/2}(m_B^2,m_{\psi}^2,M_\pi^2)}{16\pi m_B^3} \frac{(m_B^2-m_\psi^2)^2}{4 m_B m_\psi} |\tilde \alpha_1|^2 
\approx 3\, \Gamma\big(B^0 \to J/\psi \eta\big), \nonumber\\ 
\Gamma\big(B^0 \to \eta_c \pi^0\big) &= \frac{\tilde G_F^2}{4 F_\pi^2}  
\frac{\lambda^{1/2}(m_B^2,m_{\eta_c}^2,M_\pi^2) m_{\eta_c}}{16\pi m_B^2} \bigg|\tilde \alpha_2 + \frac{m_B^2+m_\psi^2}{2 m_B m_\psi} \tilde \alpha_1\bigg|^2 \nonumber\\
& \approx  \frac{|V_{cd}|^2}{2|V_{cs}|^2} \Gamma\big(B^0 \to \eta_c K^0\big) ,  \qquad 
\tilde \alpha_1=4(\alpha_1+\beta_1) , \quad 
\tilde \alpha_2=4(\alpha_2+\beta_2) ,
\end{align}
for which the branching fractions are measured~\cite{Olive:2016xmw}, 
\begin{align}
\B(B^0 \to J/\psi \pi^0) &= (1.76\pm 0.16)\times 10^{-5}, \nonumber\\
\B(B^0 \to J/\psi \eta) &= (1.08\pm 0.24)\times 10^{-5}, \nonumber\\
\B(B^0 \to \eta_c K^0) &= (8.0\pm1.2)\times 10^{-4}.
\end{align}
We therefore can fix the (combinations of) couplings $|\tilde \alpha_1| \approx 0.055\,\GeV$, which is the average of the values determined from the branching fractions into $J/\psi \pi^0$ and $J/\psi \eta$, as well as $|\tilde \alpha_2 + (m_B^2+m_\psi^2) / (2 m_B m_\psi) \cdot \tilde  \alpha_1| \approx 0.028\,\GeV$. To satisfy the latter relation, we have two choices for $|\tilde \alpha_2|$,\footnote{As a simple estimation of the error due to the unknown sign of $\tilde \alpha_1$ we focus on the linear combination of couplings $|\tilde \alpha_1 + (m_B^2+m_\psi^2) / (2 m_B m_\psi) \cdot \tilde  \alpha_2|$ entering the  $B^* \to J/\psi M_2$ amplitude, which is affected by this uncertainty on a $30\%$ level; therefore we prove the $P$-wave suppression for both values.} $|\tilde \alpha_2| \approx 0.035\,\GeV$ or $|\tilde \alpha_2| \approx 0.091\,\GeV$, with the constraint $\tilde \alpha_2/\tilde \alpha_1 < 0$.

The $B^*$-exchange graphs for $B(p_B)\to J/\psi(p_\psi) \pi^0(p_1)\eta(p_2)$ 
then lead to an amplitude contribution of the form 
{\allowdisplaybreaks
\begin{align}
\M^{\text{eff}}_{B^*} & \approx 
-\frac{\tilde G_F g \sqrt{m_B^3m_\psi}}{2 \sqrt{3} F_\pi^2}
\bigg\{  p_1^\mu \bigg[ \bigg(\tilde \alpha_1 + \frac{m_B^2 + m_\psi^2}{2m_B m_\psi} \tilde \alpha_2\bigg) \frac{1}{t-m_{B^*}^2} + \frac{1}{m_{B^*}^2} \bigg( \tilde \alpha_1 +\frac{m_B}{m_\psi} \tilde \alpha_2 \bigg)  \bigg] 
\nonumber\\
& \hspace*{1.5cm} +  p_2^\mu \bigg[ \bigg(\tilde \alpha_1 + \frac{m_B^2 + m_\psi^2}{2m_B m_\psi} \tilde \alpha_2\bigg) \frac{1}{u-m_{B^*}^2} + \frac{1}{m_{B^*}^2} \bigg( \tilde \alpha_1 +\frac{ m_B}{m_\psi} \tilde \alpha_2 \bigg)  \bigg]\nonumber\\
&\hspace*{1.5cm} + i \epsilon^{\mu\nu\alpha\beta} {p_{\psi}}_{\nu} {p_2}_\alpha {p_1}_\beta \frac{\tilde \alpha_2}{m_\psi m_B} \bigg(\frac{1}{t-m_{B^*}^2} - \frac{1}{u-m_{B^*}^2} \bigg)
\bigg\} \epsilon_\mu^*(p_\psi,\lambda) \nonumber\\
&=
-\frac{\tilde G_F g \sqrt{m_B^3m_\psi}}{4 \sqrt{3} F_\pi^2}
\bigg\{  Q^\mu \bigg[ \bigg(\tilde \alpha_1 + \frac{m_B^2 + m_\psi^2}{2m_B m_\psi} \tilde \alpha_2\bigg)\bigg(\frac{1}{t-m_{B^*}^2} - \frac{1}{u-m_{B^*}^2} \bigg) \bigg]  \nonumber\\ 
& \hspace*{0.6cm} +  P^\mu \bigg[ \bigg(\tilde \alpha_1 + \frac{m_B^2 + m_\psi^2}{2m_B m_\psi} \tilde \alpha_2\bigg) \bigg(\frac{1}{t-m_{B^*}^2} + \frac{1}{u-m_{B^*}^2} \bigg)+ \frac{2}{m_{B^*}^2} \bigg( \tilde \alpha_1 +\frac{m_B}{m_\psi} \tilde \alpha_2 \bigg)  \bigg] 
\nonumber\\
&\hspace*{0.6cm}  + i \epsilon^{\mu\nu\alpha\beta} {p_{\psi}}_{\nu} {P}_\alpha {Q}_\beta \frac{\tilde \alpha_2}{m_\psi m_B} \bigg(\frac{1}{t-m_{B^*}^2} - \frac{1}{u-m_{B^*}^2} \bigg)
\bigg\} \epsilon_\mu^*(p_\psi,\lambda) ,
\end{align}}%
where terms of order $M_{\pi/\eta}^2/m_B^2$ or $(m_{B^*}-m_B)/m_B$ have been neglected.
In order to project this expression onto the transversity form factors $\F_0$, $\F_\parallel$, we need to replace
$Q^\mu$ by $Q_{(\parallel)}^\mu = Q^\mu + \gamma P^\mu + \ldots$, 
where $\gamma$ can be read off from 
eq.~\eqref{eq:newbasis}.  We find that we can approximate $\gamma$ according to
\begin{equation}
\gamma = 1 + \frac{2(t-m_B^2)}{m_B^2-m_\psi^2} + \Op\big(s,M_\pi^2,M_\eta^2\big)
= -1 - \frac{2(u-m_B^2)}{m_B^2-m_\psi^2} + \Op\big(s,M_\pi^2,M_\eta^2\big),
\end{equation}
such that
{\allowdisplaybreaks
\begin{align}
\F_0^{(B^*)}(s,t,u) &=\frac{ g \sqrt{m_Bm_\psi} \,X}{\sqrt{3} F_\pi^2 } \frac{1}{m_B^2-m_{\psi}^2} 
\bigg( \tilde \alpha_2 + \frac{m_B^2+m_\psi^2}{2 m_B m_\psi} \tilde \alpha_1  \bigg), \nonumber\\ 
\F_\parallel^{(B^*)}(s,t,u) &= -\frac{g \,\sqrt{m_B^3m_\psi \,s}}{4 \sqrt{3} F_\pi^2} \bigg(\tilde \alpha_1 + \frac{m_B^2 + m_\psi^2}{2m_B m_\psi} \tilde \alpha_2\bigg) 
\bigg[\frac{1}{t-m_{B^*}^2} - \frac{1}{u-m_{B^*}^2} \bigg], \nonumber\\ 
\F_\perp^{(B^*)}(s,t,u) &= \frac{g \sqrt{m_B \,s} \,X}{4 \sqrt{3m_\psi} F_\pi^2} \,\tilde \alpha_2 
\bigg[\frac{1}{t-m_{B^*}^2} - \frac{1}{u-m_{B^*}^2} \bigg] . \label{eq:F0pp}
\end{align}}%
The partial-wave expansion of  $\F_0^{(B^*)}$ contains an $S$-wave only.
We find that the $S$-wave expression induced by $B^*$-exchange does not actually include
a left-hand cut: the $t$- and $u$-channel pole contributions cancel, leaving behind an effectively
point-like source for an $S$-wave $\pi\eta$ pair. 
In this way, $B^*$-exchange provides a model for the coupling constant that was obtained
purely phenomenologically by fitting to data in the main text.
In order to compare the strength of the $B^*$-exchange-induced $S$-wave to the phenomenological one, we calculate the analogue of the (fitted and afterwards properly normalized) constant $\bar b_0^n \approx 2.8 \cdot 10^{-10}\,\GeV^{-3}$ by means of eq.~\eqref{eq:F0s}.  We find a strength $|\bar b_0^{B^*}| \approx 1.6 \cdot 10^{-10}\,\GeV^{-3}$: the combination of the two couplings $\tilde \alpha_{1/2}$ is exactly the one fixed from $B^0\to \eta_c K^0$ above, hence the ambiguity in $\tilde\alpha_2$ just translates into a sign ambiguity once more.
The effective coupling strength therefore indeed produces an $S$-wave rate of the correct order of magnitude, pointing towards an essential role of the $B^*$ in the explanation of the production mechanism; a more systematic investigation of this strength is beyond the scope of the present article.

Our main focus here is rather on the $P$-waves 
in $\F_\parallel^{(B^*)}$ and $\F_\perp^{(B^*)}$, which are given by 
{\allowdisplaybreaks
\begin{align}
\F_\parallel^{(B^*)(P)}(s) &= 
-\frac{g \,\sqrt{m_B^3m_\psi \,s}}{4 \sqrt{2} F_\pi^2} \bigg(\tilde \alpha_1 + \frac{m_B^2 + m_\psi^2}{2m_B m_\psi} \tilde \alpha_2\bigg) \,
 \frac{1}{XY} 
\bigg( v_+ - \frac{v_+^2-1}{2}\log\frac{v_++1}{v_+-1} \nonumber\\
& \quad -v_- + \frac{v_-^2-1}{2}\log\frac{v_-+1}{v_--1} \bigg) , \nonumber\\
\F_\perp^{(B^*)(P)}(s) &= 
\frac{g \sqrt{m_B \,s}}{4 \sqrt{2m_\psi} F_\pi^2} \frac{\tilde \alpha_2}{Y} 
\bigg( v_+ - \frac{v_+^2-1}{2}\log\frac{v_++1}{v_+-1} -v_- + \frac{v_-^2-1}{2}\log\frac{v_-+1}{v_--1} \bigg) , \nonumber\\
v_\pm &= \frac{1}{2XY}\bigg[ \Sigma -2m_{B^*}^2 -s\pm\frac{\Delta_m}{s}\bigg], 
\end{align}}%
and the pairs of terms depending on $v_\pm$, coming from the $t$- and $u$-channel pole terms, 
almost cancel each other (such that the $D$-waves are actually as large as the $P$-waves around $\sqrt{s}\approx 1\,\GeV$).
Both yield contributions to $\diff\Gamma/\diff\sqrt{s}$ suppressed relative to $\F_0^{(B^*)}$ by two orders of magnitude,
proving yet again the strong dominance of the $S$-wave in this decay.

\bibliography{BJpsi}


\end{document}